\documentclass[]{spie}

\newcommand{\Grad}{{\boldsymbol{\nabla}}}
\newcommand{\Div}{\Grad\cdot}

\newcommand{\infd}{\text{d}}

\newcommand{\td}[2]{\frac{\text{d}{#1}}{\text{d}{#2}}}
\newcommand{\tdtwo}[2]{\frac{\text{d}^2{#1}}{\text{d}{#2}^2}}

\newcommand{\pd}[2]{\frac{\partial{#1}}{\partial{#2}}}
\newcommand{\pdtwo}[2]{\frac{\partial^2{#1}}{\partial{#2}^2}}

\newcommand{\trace}[1]{\text{tr}\left({#1}\right)}
\newcommand{\e}{\textrm{e}}

\newcommand{\I}{\mathbf{I}}

\newcommand{\segflux}{\boldsymbol{q}}

\newcommand{\displace}{\boldsymbol{u}}
\newcommand{\vel}{\boldsymbol{v}}

\newcommand{\stressten}{\boldsymbol{\sigma}}
\newcommand{\strainten}{\boldsymbol{e}}

\newcommand*{\rom}[1]{\expandafter\@slowromancap\romannumeral #1@}

\usepackage[OT2,T1]{fontenc}
\DeclareSymbolFont{cyrletters}{OT2}{wncyr}{m}{n}
\DeclareMathSymbol{\Sha}{\mathalpha}{cyrletters}{"58}
\usepackage{mathtools}
\usepackage{amsmath,amsfonts,amssymb,esint}
\usepackage{graphicx}
\usepackage{lipsum}
\usepackage{natbib}
\usepackage[colorlinks=true, allcolors=blue]{hyperref}
\usepackage{multicol}
\usepackage{subcaption}
\usepackage{tocbibind} 
\usepackage[toc,page]{appendix} 

\usepackage{setspace}
\usepackage{lineno}

\title{Episodic, compression-driven fluid venting in layered sedimentary basins}
\author[1]{Luke~M.~Kearney}
\author[2]{Christopher~W.~MacMinn}
\author[1]{Richard~F.~Katz}
\author[1]{Chris~Kirkham}
\author[1]{Joe~Cartwright}

\affil[1]{\normalsize{\textit{Department of Earth Sciences, University of Oxford, Oxford OX1 3AN, United Kingdom}}}
\affil[2]{\normalsize{\textit{Department of Engineering Science, University of Oxford, Oxford OX1 3PJ, United Kingdom}}}

\begin{document}

\maketitle
\begin{abstract}
Fluid venting phenomena are prevalent in sedimentary basins globally. Offshore, these localised fluid-expulsion events are archived in the geologic record by the resulting pockmarks at the sea-floor. Venting is widely interpreted to occur via hydraulic fracturing, which requires near-lithostatic pore pressures for initiation. One common driver for these extreme pressures is horizontal tectonic compression, which pressurises the entire sedimentary column over a wide region. Fluid expulsion leads to a sudden, local relief of this pressure, which then gradually recharges through continued compression, leading to episodic venting. Pressure recharge will also occur through pressure diffusion from neighbouring regions that remain pressurised, but the combined role of compression and pressure diffusion in episodic venting has not previously been considered. Here, we develop a novel poroelastic model for episodic, compression-driven venting. We show that compression and pressure diffusion together set the resulting venting period. We derive a simple analytical expression for this venting period, demonstrating that pressure diffusion can significantly reduce the venting period associated with a given rate of compression. Our expression allows this rate of compression to be inferred from observations of episodic venting. We conclude that pressure diffusion is a major contributor to episodic fluid venting in mudstone-dominated basins.
\end{abstract} 


\vspace{1cm}

\section{Introduction}
\noindent

Fluid venting phenomena have been frequently observed in sedimentary basins since the advent of 3D seismic imaging \citep{loseth20011000, hovland1988seabed}. The vents themselves are localised, typically comprising cylindrical conduits known as fluid-escape pipes that can create pockmarks or feed effusive mud volcanoes \citep{huuse2010subsurface, cartwright2015seismic}. Venting is thought to initiate when the seal of a pressurised reservoir fails through hydraulic fracturing, creating a high-permeability pathway for the transport of basinal fluids through kilometres of low-permeability rock \citep{cartwright2021quantitative}. This mode of seal failure poses clear risks for the subsurface storage of hydrogen and the long-term sequestration of anthropogenic waste such as carbon dioxide (CO$_2$). Indeed, unexpected vertical fluid migration at the Sleipner CO$_2$ storage pilot site is likely due to exploitation of pre-existing conduits \citep{arts2004seismic, cavanagh2014sleipner}.

Repeated fluid venting from a fixed locus has been documented in a subset of cases \citep{deville2010fluid, andresen2011bulls, cartwright2018direct, oppo2021leaky, kirkham2022episodic}. In each of these cases, venting occurs in discrete episodes of fluid expulsion separated by long quiescent periods. In the North Levant Basin, located in the Eastern Mediterranean, the presence of a flowing salt sheet enables dating of individual venting episodes \citep{oppo2021leaky, evans2020taking, cartwright2021quantitative}, thus providing a robust basis for investigating episodic fluid venting. More than 300 fluid escape pipes record episodic venting through this $\sim$1.5~km-thick layer of low-permeability salt. The salt overlies a $\sim$3~km-thick clastic succession dominated by mudstone. These fluid-escape pipes are interpreted to form vertically from the crests of folded sandstone reservoirs, terminating at the seafloor as pockmarks. Viscous flow of the salt layer deforms the relic pipes over geological time, such that repeated venting leads to a linear trail of pockmarks along the direction of salt flow \citep{cartwright2018direct}. Thirteen pockmark trails have been observed across the North Levant Basin, each recording up to 45 venting episodes since $\sim$2~Ma \citep{oppo2021leaky}. Dating of these venting episodes reveals a typical time interval between episodes (i.e.,~venting period) of $\sim$100~kyr \citep{oppo2021leaky, evans2020taking, cartwright2021quantitative}.

The initiation of a vent via hydraulic fracturing requires fluid pressure in excess of the minimum horizontal compressive stress \citep{price1990analysis, scandella2011conduit}. Once initiated, venting continues until this overpressure is sufficiently relieved that the pathway closes. Subsequently, during quiescence, fractures may self-heal by solid creep, swelling, and mineral precipitation \citep{bock2010self, chen2013self}. In the North Levant Basin, previous pathways are deformed and advected away from their original trajectory by salt flow. Hence, episodic venting requires the repeated recharge of overpressure to the original point of failure, implying that the overpressure mechanism remains active across venting episodes. The disparity between the rapid drop in pressure during venting and the slow growth of pressure during recharge suggests that the time-history of reservoir pressure across multiple episodes resembles a sawtooth pattern, with the up-slope representing the rate of pressure recharge and the amplitude representing the pressure drop during venting \citep{cartwright2021quantitative}. \cite{cartwright2021quantitative} attribute this pressure drop to be the tensile strength of the sealing rock, estimated to range from 0.6~MPa to 2~MPa. Using the sawtooth concept and the measured period between venting episodes, \cite{cartwright2021quantitative} inferred a rate of pressure recharge in the North Levant Basin of $\sim$9~MPa/Myr.

Overpressure can be generated by various mechanisms \citep{osborne1997mechanisms}. For the Levant basin, \cite{cartwright2021quantitative} ascribe overpressure generation to regional tectonic compression on the basis of qualitative physical arguments. Previous studies have used numerical models to predict overpressures due to tectonic compression and quantify the role of factors such as duration and rate of shortening \citep{luo2004quantitative, obradors2017assessing, obradors2017hydromechanical, maghous2014two}. For example, \cite{obradors2017assessing} showed that an overpressure of $\sim$10~MPa can be generated by a shortening of $\sim$10\% over a period of $\sim$100~kyr. However, overpressure will typically be heterogeneously distributed throughout the sedimentary column. \cite{ge1992hydromechanical} showed that tectonic compression pressurises stratigraphic layers at different rates due to their different elastic properties. These pressure differences equilibrate over time through vertical fluid redistribution, which can be described mathematically as the diffusion of pressure.

Pressure diffusion between sedimentary layers has been investigated in many previous works \citep[e.g.,][and refs.~therein]{muggeridge2004dissipation, muggeridge2005rate}. The primary concern of these studies has been to estimate the timescales and mechanisms of pressure redistribution through low-permeability layers. As a result, these studies typically focus on the diffusive equilibration of an initially non-hydrostatic pressure distribution while neglecting the origin of that distribution or any ongoing sources of overpressure generation. This omission may not always be justified, given that mechanisms such as tectonic compression persist for millions of years and are physically independent of pressure redistribution. Moreover, these studies generally neglect punctuated effects that modify the pressure, such as venting. An exception is \cite{luo2016overpressure}, who investigated mechanisms of pressure dissipation including hydraulic fracturing.

Despite the large body of relevant work, most previous studies have neglected at least one of the three key components of episodic venting: pressure build-up, pressure diffusion, and hydraulic fracturing. The studies that include all three components predict episodic venting. However, these models incorporate a variety of additional physics such as reaction and heat transport, necessitating numerical solution \citep{dewers1994nonlinear, l2000simple}. The complexity and computational expense of these models limits them to generating a small set of results for a specific setting, making it difficult to develop more general insight. Such insight is facilitated by a simplified theory that incorporates only the physical processes needed to describe the general, episodic dynamics of fluid venting in sedimentary basins. Moreover, measurements of the venting period are readily interpreted in this analytical context.

Here we develop a poroelastic model of tectonic overpressure generation, diffusive pressure redistribution, and fluid venting in layered sedimentary basins. We derive analytical solutions that elucidate the associated pressure dynamics and the parametric controls on venting. We show in particular that the venting period $\tau$ is given by $\tau \propto (\Delta P/\dot{e}_{xx}) / (1+\nu/\gamma)$, where $\Delta P$ is the pressure drop from each venting event, $\dot{e}_{xx}$ is the horizontal strain rate due to tectonic compression and $\nu$ and $\gamma$ are dimensionless parameters that are defined below. The quantity $\Delta P/\dot{e}_{xx}$ is proportional to the venting period in the absence of pressure diffusion, as estimated by \citet{cartwright2021quantitative}. We refer to the dimensionless quantity $(1+\nu/\gamma)$ as the \textit{venting frequency multiplier} because it reduces the venting period relative to the compression-only case. We show that this frequency multiplier can be estimated using the thickness ratio of the mudstone and sandstone layers. In mudstone-dominated basins where fluid venting phenomena are commonly observed \citep{cartwright2015seismic}, pressure recharge and venting period are controlled by pressure diffusion.

The remainder of the manuscript is organised as follows. In \S\ref{sec:compression}, we derive and solve the poroelastic equations governing tectonic compression of, and pressure diffusion between, sedimentary layers in the absence of fluid venting. In \S\ref{sec:venting}, we explore the response of the system to fluid venting without compression. In \S\ref{sec:episodic}, we combine solutions from \S\ref{sec:compression} and \S\ref{sec:venting} to obtain a full model for episodic venting; we then derive analytical solutions for periodic venting. In \S\ref{sec:discussion}, we discuss the wider implications of this work, as well as limitations and potential generalisations of the model. In \S\ref{sec:conclusions}, we conclude with a summary and suggestions for promising avenues of future work.
\section{Model}
\label{sec:model}

\subsection{Compression}
\label{sec:compression}

\begin{figure}[!b]
\centering
\begin{subfigure}[b]{0.99\textwidth}
   \includegraphics[width=1\linewidth]{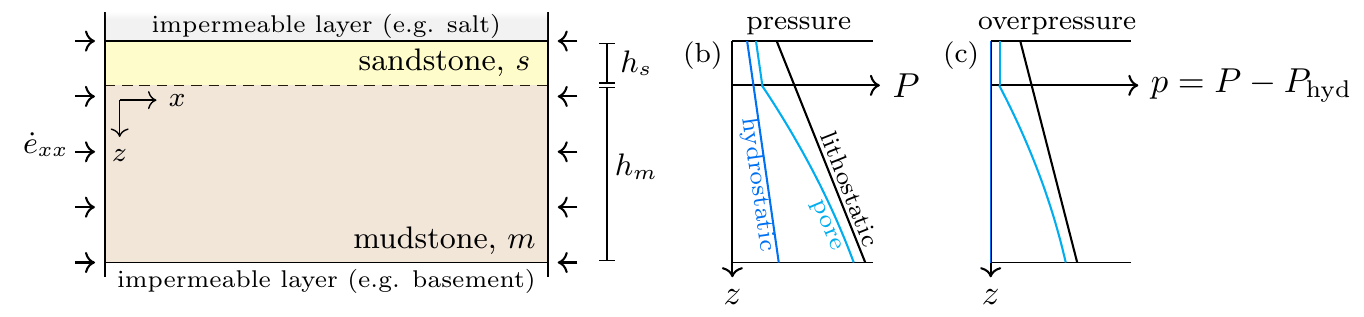}
\end{subfigure}
\caption{Schematic cross-section of sedimentary-basin model. (a) We consider the tectonic compression of two permeable sedimentary layers, sandstone (yellow) and mudstone (brown), at a constant horizontal strain rate $-\dot e_{xx}$. The system is sealed by impermeable layers above the sandstone and below the mudstone. (b) Example pressure--depth plot showing the hydrostatic pressure (blue), pore pressure resulting from compression (cyan) and lithostatic stress (black). (c) As in (b), but now showing the lithostatic stress in excess of hydrostatic (black) and the pore overpressure $p$ (cyan), which is the pressure in excess of hydrostatic $P_{\mathrm{hyd}}$.
\label{fig:model}}
\end{figure}

We consider two horizontal layers of rock, a sandstone with thickness $h_s$ overlying a mudstone with thickness $h_m$, as illustrated in Figure~\ref{fig:model}a. To focus on large-scale pressurisation from regional tectonic compression, we assume that these layers have a large lateral extent, such that pressure diffusion occurs exclusively through vertical fluid migration. The development of large overpressure from tectonic compression thus requires that compression be rapid relative to pressure diffusion and/or that vertical flow be obstructed. 

Sedimentary basins are typically underlain by dense basement rock, so we assume the existence of an impermeable layer below the mudstone. Furthermore, motivated by sedimentary basins such as the Levant basin that are capped with an extensive, thick salt layer, we apply the same assumption above the sandstone. Salt is considered to be impermeable on geological timescales \cite[though see][]{ghanbarzadeh2015deformation}. This model configuration prohibits vertical pressure diffusion across the salt, but allows for sudden fluid expulsion via hydraulic fracturing. The theory below could be generalised to allow for a `seal' with a small but nonzero permeability. 

As in previous studies that consider pressure diffusion between sedimentary layers, we assume that flow is single-phase, isothermal and one-dimensional \citep{bredehoeft1968maintenance, neuzil1986groundwater, luo1997sealing, muggeridge2004dissipation}. Crucially, we deviate from these previous studies by modelling the evolution of pressure due to ongoing (rather than historical) tectonic compression. Tectonic compression has been conceptualised as a bulldozer imparting sufficient differential stress to deform weaker sediments \citep{byrne1993mechanical}. It has been modelled mathematically as an imposed, constant horizontal strain rate; strain rates are routinely used to quantify tectonic deformation \citep[e.g.][]{kahle1998strain, kreemer2014geodetic}. We denote the imposed strain rate as $-\dot e_{xx}$, such that a positive value of $\dot e_{xx}$ indicates shortening. In the absence of venting, tectonic shortening is accommodated through compression of the pore fluid and/or of the solid grains. We refer to this specific process as tectonic compression. 

Assuming that the solid skeleton obeys linear elasticity and adopting the sign convention that tension is positive, the effective stress tensor $\stressten'$ is related to the strain tensor $\strainten$ via
\begin{equation}
\label{eqn:constitutive}
    \stressten' = \lambda \trace \strainten \I + 2 \mu \strainten,
\end{equation}
where $\lambda$ and $\mu$ are the drained Lam\'e parameters and $\strainten \equiv \tfrac{1}{2}[\Grad \displace + (\Grad \displace)^\mathrm{T}]$ with $\displace$ denoting the solid displacement. The effective stress is related to the total stress $\stressten$ and pore pressure $P$ by Terzaghi's principle,
\begin{equation}
\label{eqn:terzaghi}
    \stressten = \stressten' - \alpha P \I,
\end{equation}
where $\alpha$ is Biot's coefficient. Mass conservation leads to the storage equation \citep[][see Supplementary Material~\ref{sup:storage}]{verruijt1969elastic}, which is equivalent to the classical continuity equation presented by \cite{biot1941general},
\begin{equation}
\label{eqn:storage}
    \alpha \pd{e}{t} + S \pd{P}{t} = - \Div \segflux,
\end{equation}
where $e \equiv \trace \strainten$ is the volumetric strain, $\segflux$ is the Darcy flux of fluid through the pore space and $S \equiv \phi c_\ell + (\alpha - \phi) c_g$ is the storativity, with porosity $\phi$, fluid compressibility $c_\ell$ and grain compressibility $c_g$. The time derivative of the $xx$-component of Equation~(\ref{eqn:constitutive}) implies that
\begin{equation}
    \pd{\sigma_{xx}'}{t} = \lambda \pd{e}{t} - 2 \mu \dot e_{xx},
\end{equation}
and the trace of Equation~(\ref{eqn:constitutive}) implies that $(3\lambda + 2\mu) \, e = \trace{\stressten'}$. From these results and Equation~(\ref{eqn:terzaghi}), we arrive at
\begin{equation}
\label{eqn:5}
    \pd{P}{t} = \frac{\lambda + \mu}{\alpha}\pd{e}{t} + \frac{\mu}{\alpha} \dot e_{xx} - \frac{1}{2\alpha}\bigg( \pd{\sigma_{yy}}{t} + \pd{\sigma_{zz}}{t} \bigg).
\end{equation}
Equation~(\ref{eqn:5}) describes the evolution of pore pressure in response to changes in strain and total stress. The total vertical stress at a fixed depth can increase ($\partial_t \sigma_{zz} < 0$) in response to folding and thrust faulting, which is a common consequence of tectonic shortening. In the North Levant Basin, folding and thrusting leads to localised salt thinning, resulting in a slowly decreasing total vertical stress \citep{cartwright2021quantitative}. For simplicity, we neglect this minor effect by assuming that $\partial_t \sigma_{zz} = 0$. The evolution of the total horizontal stress in the orthogonal direction, $\partial_t \sigma_{yy}$ is less clear. Two endmember assumptions are that the orthogonal total stress remains constant ($\partial_t \sigma_{yy} = 0$), or that the associated strain remains constant ($\partial_t e_{yy} = 0$). The former unconditionally allows for hydraulic fracturing, whereas the latter may not in some cases (Supplementary Material~\ref{sup:hydfrac}). For simplicity, we proceed with the former assumption and take $\sigma_{yy}$ to be constant. Combining these assumptions with Eqs.~(\ref{eqn:storage}) and (\ref{eqn:5}) gives
\begin{equation}
\label{eqn:funda}
    \pd{P}{t} = \frac{\alpha \mu \dot e_{xx}}{\alpha^2 + S(\lambda + \mu)} - \frac{\lambda + \mu}{\alpha^2 + S(\lambda + \mu)} \Div \segflux.
\end{equation}
Thus, two processes drive changes in pore pressure: compression and fluid flow. The first term in Eq.~(\ref{eqn:funda}) corresponds to compression, which acts to increase pressure everywhere at a rate determined by the compression rate and the poroelastic properties of the medium. The second term in Eq.~(\ref{eqn:funda}) demonstrates that pressure increase at a point is impeded by a net export of fluid ($\Div \segflux > 0$) or enhanced by a net import ($\Div \segflux < 0$). 

For a system that is both laterally extensive and laterally homogeneous (i.e., no variations in $x$ or $y$; Fig.~\ref{fig:model}), fluid flow is limited to the vertical direction, $\segflux \equiv q \, \hat{\mathbf{z}}$. Then, applying Eq.~(\ref{eqn:funda}) to each layer,
\begin{subequations}
\label{eqn:layers}
\begin{align}
    \displaystyle\pd{P_s}{t} = \frac{\alpha_s \mu_s \dot e_{xx}}{\alpha_s^2 + S_s(\lambda_s + \mu_s)} - \frac{\lambda_s + \mu_s}{\alpha_s^2 + S_s(\lambda_s + \mu_s)} \pd{q}{z} \hspace{0.5cm} &\mathrm{for} \,\, z \in [-h_s, 0], \\
    \displaystyle\pd{P_m}{t} = \frac{\alpha_m \mu_m \dot e_{xx}}{\alpha_m^2 + S_m(\lambda_m + \mu_m)} - \frac{\lambda_m + \mu_m}{\alpha_m^2 + S_m(\lambda_m + \mu_m)} \pd{q}{z} \hspace{0.5cm} &\mathrm{for} \,\, z \in [0, h_m].
\end{align}
\end{subequations}
where the subscripts $s$ and $m$ represent properties of the sandstone and of the mudstone, respectively. In this system, the hydrostatic contribution to the pressure remains constant. It has no effect on the dynamics and hence we replace total pressure $P$ with overpressure $p$. The overpressure is the pressure in excess of hydrostatic, $p \equiv P-P_{\mathrm{hyd}}$ (see Fig.~\ref{fig:model}b,c),
\begin{subequations}
\label{eqn:layers}
\begin{align}
    \label{eqn:sandlayer}
    \displaystyle\pd{p_s}{t} = \frac{\alpha_s \mu_s \dot e_{xx}}{\alpha_s^2 + S_s(\lambda_s + \mu_s)} - \frac{\lambda_s + \mu_s}{\alpha_s^2 + S_s(\lambda_s + \mu_s)} \pd{q}{z} \hspace{0.5cm} &\mathrm{for} \,\, z \in [-h_s, 0], \\
    \label{eqn:mudlayer}
    \displaystyle\pd{p_m}{t} = \frac{\alpha_m \mu_m \dot e_{xx}}{\alpha_m^2 + S_m(\lambda_m + \mu_m)} - \frac{\lambda_m + \mu_m}{\alpha_m^2 + S_m(\lambda_m + \mu_m)} \pd{q}{z} \hspace{0.5cm} &\mathrm{for} \,\, z \in [0, h_m].
\end{align}
\end{subequations}
Sandstones typically have permeabilities that are many orders of magnitude larger than those of mudstones. Consequently, pressure diffuses much faster in sandstone than in mudstone. Hence over timescales of pressure diffusion in the mudstone, the overpressure in the sandstone is approximately vertically uniform. Considering this, we integrate Equation (\ref{eqn:sandlayer}) over the thickness of the sandstone,
\begin{equation}
    \td{\overline{p_s}}{t} = \frac{\alpha_s \mu_s \dot e_{xx}}{\alpha_s^2 + S_s(\lambda_s + \mu_s)} - \frac{\lambda_s + \mu_s}{\alpha_s^2 + S_s(\lambda_s + \mu_s)} \frac{q(0,t) - q(-h_s, t)}{h_s},
\end{equation}
where $\overline{p_s}$ is the depth-averaged overpressure in the sandstone. Therefore, the rate of change of the average overpressure in the sandstone is given by a term from compression and a term from the difference in flux across the boundaries. The sandstone is overlain by an impermeable layer so $q(-h_s, t) = 0$. Since the overpressure in the sandstone is approximately uniform, we assume $p_s = \overline{p_s}$ and obtain
\begin{equation}
    \label{eqn:sand}
    \td{p_s}{t} = \frac{\alpha_s \mu_s \dot e_{xx}}{\alpha_s^2 + S_s(\lambda_s + \mu_s)} - \frac{\lambda_s + \mu_s}{\alpha_s^2 + S_s(\lambda_s + \mu_s)} \frac{q(0,t)}{h_s},
\end{equation}
where the flux out of the sandstone $q(0,t)$ must be equal to the flux into the top of the mudstone (in the absence of venting). Fluid transport in the mudstone is governed by Darcy's law,
\begin{equation}
\label{eqn:darcy}
    q = -\frac{k_m}{\eta} \pd{p_m}{z},
\end{equation}
where $\eta$ is the viscosity of the fluid and $k_m$ is the permeability of the mudstone, both assumed to be constant. Equations (\ref{eqn:mudlayer}), (\ref{eqn:sand}) and (\ref{eqn:darcy}) combine to form a coupled system for the overpressure of each layer,
\begin{subequations}
\label{eqn:11}
\begin{align}
\label{eqn:11a}
    \td{p_s}{t} = \frac{\alpha_s \mu_s \dot e_{xx}}{\alpha_s^2 + S_s(\lambda_s + \mu_s)} + \frac{\lambda_s + \mu_s}{\alpha_s^2 + S_s(\lambda_s + \mu_s)} \frac{1}{h_s}\frac{k_m}{\eta}\pd{p_m}{z}\bigg|_{z=0} \hspace{0.5cm} &\mathrm{at} \,\, z = 0, \\
\label{eqn:11b}
    \pd{p_m}{t} = \frac{\alpha_m \mu_m \dot e_{xx}}{\alpha_m^2 + S_m(\lambda_m + \mu_m)} + \frac{\lambda_m + \mu_m}{\alpha_m^2 + S_m(\lambda_m + \mu_m)} \frac{k_m}{\eta} \pdtwo{p_m}{z} \hspace{0.5cm} &\mathrm{for} \,\, z \in [0, h_m].
\end{align}
\end{subequations}
Equation (\ref{eqn:11a}) is an ordinary differential equation for the time-evolution of the overpressure in the sandstone. Equation~(\ref{eqn:11b}) is a partial differential equation for the overpressure in the mudstone in depth and time. The latter requires two boundary conditions. The first boundary condition is that the sandstone and mudstone overpressures must match at the contact, $p_m(0, t) = p_s$. The second boundary condition is that there is no fluid flux through the impermeable layer at the bottom of the mudstone $\partial p_m/\partial z |_{h_m} = 0$. Equations~(\ref{eqn:11}) are simplified by introducing the following parameters,
\begin{equation}
\label{eqn:ddef}
    D_m = \frac{k_m}{\eta} \frac{\lambda_m + \mu_m}{\alpha_m^2 + S_m(\lambda_m + \mu_m)}, \hspace{1cm} D_s = \frac{k_m}{\eta}\frac{\lambda_s + \mu_s}{ \alpha_s^2 + S_s(\lambda_s + \mu_s)},
\end{equation}
\begin{equation}
\label{eqn:gammadef}
    \Gamma_m = \frac{\alpha_m \mu_m \dot e_{xx}}{\alpha_m^2 + S_m(\lambda_m + \mu_m)}, \hspace{1cm} \Gamma_s = \frac{\alpha_s \mu_s \dot e_{xx}}{\alpha_s^2 + S_s(\lambda_s + \mu_s)},
\end{equation}
The parameters $\Gamma_m$ and $\Gamma_s$ represent the rates of pressure build-up in the mudstone and sandstone layers, respectively, due to compression. The parameter $D_m$ represents the diffusivity of pressure across the mudstone whereas the parameter $D_s$ represents the diffusivity of pressure across the sandstone--mudstone boundary. The full model can then be written,
\begin{equation}
\label{eqn:gov1}
    \pd{p_m}{t} = \Gamma_m + D_m \pdtwo{p_m}{z} \hspace{0.5cm} \mathrm{for} \,\, z \in [0, h_m],
\end{equation}
with boundary conditions,
\begin{equation}
\label{eqn:gov2}
    \begin{rcases}
        \displaystyle\pd{p_m}{t} = \Gamma_s + \frac{D_s}{h_s}\pd{p_m}{z} \hspace{0.5cm} &\mathrm{at} \,\, z=0, \\
        \displaystyle\pd{p_m}{z} = 0 \hspace{0.5cm} &\mathrm{at} \,\, z=h_m,
    \end{rcases}
\end{equation}
where $p_s = p_m(0, t)$. We finally assume that the mudstone and sandstone are initially at hydrostatic pressure,
\begin{equation}
    p_m(z, 0) = 0.
\end{equation}
This initial condition is chosen for simplicity and is not essential to the model; in many circumstances, it may be more appropriate to invoke a different initial pressure distribution. Here, we are interested in the general response to tectonic compression, decoupled from the basin-specific response due to an initial pressure disequilibrium.

\subsubsection{Nondimensionalised equations}

We nondimensionalise this system of equations using mudstone properties, introducing a dimensionless depth $z^*$, mudstone pressure $p_m^*$, and time $t^*$ defined as
\begin{equation}
    z \equiv h_m z^*, \hspace{1cm} p_m \equiv \dfrac{\Gamma_m h_m^2}{D_m} p_m^*, \hspace{1cm} t \equiv \dfrac{h_m^2}{D_m} t^*.
\end{equation}
Note that time is scaled by the characteristic time for a pressure perturbation to diffuse across the mudstone. This process typically requires thousands to millions of years and thus we scale time to study the pressure behaviour over these geological timescales. Pressure is scaled by the increase in mudstone pressure due to compression over this characteristic time. This formulation helps to highlight the effects of differences in layer properties on pressure behaviour. The dimensionless equation for the mudstone is then
\begin{equation}
\label{eqn:nondiminternal}
    \pd{p_m^*}{t^*} = 1 + \pdtwo{p_m^*}{z^*} \hspace{0.5cm} \mathrm{for} \,\, z^* \in [0, 1].
\end{equation}
with boundary conditions,
\begin{equation}
\label{eqn:nondimbdy}
    \begin{rcases}
        \displaystyle\pd{p_m^*}{t^*} = \gamma + \nu \pd{p_m^*}{z^*} \hspace{0.5cm} &\mathrm{at} \,\, z^*=0, \\
        \displaystyle\pd{p_m^*}{z^*} = 0 \hspace{0.5cm} &\mathrm{at} \,\, z^*=1,
    \end{rcases}
\end{equation}
and initial condition $p_m^*(z^*, 0) = 0$. Two dimensionless parameters $\gamma$ and $\nu$ emerge to characterise the system,
\begin{equation}
\label{eqn:nugammadef}
    \gamma \equiv \dfrac{\Gamma_s}{\Gamma_m}, \hspace{1cm} \nu \equiv \dfrac{D_s}{D_m}\dfrac{h_m}{h_s}.
\end{equation}
The parameter $\gamma$ measures the rate of pressure build-up of the sandstone relative to that of the mudstone, which depends on the material properties of each rock type and is independent of strain rate. We refer to the parameter $\nu$ as the hydraulic capacitance ratio. The hydraulic capacitance of a layer is given by the product of compressibility and volume (or thickness, in one dimension). Hydraulic capacitance measures the change in bulk volume (or pore volume, in the absence of venting) associated with a unit change in the pressure of that layer. If the sandstone layer is thinner and less compressible than the mudstone, as would typically be expected, then the sandstone will have a lower hydraulic capacitance than the mudstone ($\nu > 1$). As a result, a transfer of fluid from the mudstone to the sandstone would lead to a pressure decrease in the mudstone and a pressure increase in the sandstone, but the increase would be larger than the decrease by a factor of $\nu$. This asymmetry is central to our results here because it implies that a pressurised mudstone can fully recharge a sandstone that has been depressurised by venting $\nu$ times, even without further tectonic compression. Here we note that a low-porosity mudstone may have insufficient pore volume to recharge a high-porosity sandstone. We neglect this effect, however, because in many cases, highly overpressured mudstones are underconsolidated, with porosities comparable to typical sandstones.

\begin{figure}[!htbp]
\centering
\begin{subfigure}[b]{0.8\textwidth}
   \includegraphics[width=1\linewidth]{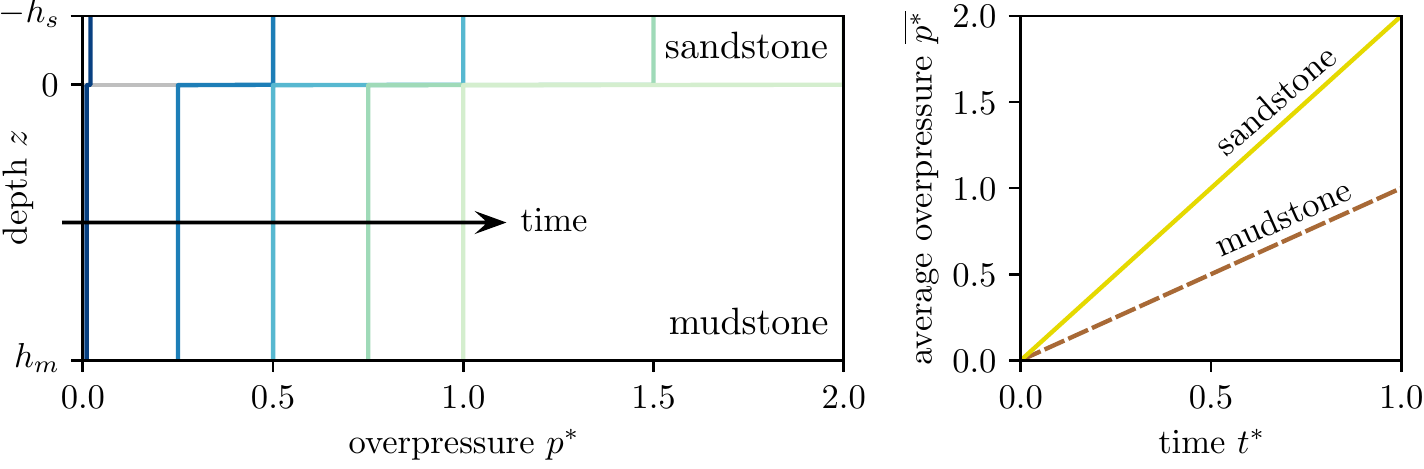}
\end{subfigure}
\caption{Early-time compression solution for $\gamma = 2$. Left: Overpressure versus depth at times $t^* = (0.01, 0.25, 0.5, 0.75, 1)$ (dark to light colours). Right: Time evolution of the average overpressure in the sandstone (solid yellow) and mudstone (dashed brown). \label{fig:earlytimes}}
\end{figure}

We solve the system of equations (\ref{eqn:nondiminternal}) and (\ref{eqn:nondimbdy}) via Laplace transforms (Supplementary Material~\ref{sup:compsol}) to arrive at
\begin{equation}
\label{eqn:compsol}
    p_m^* = t^* + (\gamma-1) \bigg\{ \frac{t^*}{1+\nu} + 2 \nu \sum_{j=1}^\infty \frac{ \cos \{\xi_j (1-z^*)\} }{\cos\xi_j}\frac{1-\exp(-\xi_j^2 t^*) }{\xi_j^2 (\nu^2 + \nu + \xi_j^2 )}\bigg\},
\end{equation}
where $\xi_j$ is the $j^{\mathrm{th}}$ solution to $\tan\xi_j = -\xi_j/\nu$. The first term $t^*$ is the pressure response to the compression of the mudstone. The remaining terms express pressure communication between the sandstone and mudstone. Two time-regimes emerge because compression begins to act instantaneously whereas diffusion takes effect over a characteristic timescale $h_m^2/D_m$. Our dimensionless time variable $t^*$ is scaled by the latter and so diffusion becomes important for dimensionless times $t^* \sim 1$ and larger. For early times $t^* \ll 1$, diffusion is negligible and layer pressures increase independently. It can be shown that this early-time behaviour is given by (Supplementary Material~\ref{sup:compearly})
\begin{equation}
    p_m^* \sim t^*, \hspace{1.5cm} p_s^* \sim \gamma t^*, \hspace{1.5cm} \mathrm{for} \,\, t^* \ll 1.
\end{equation}
In this limit, the mudstone pressure is uniform in space. The mudstone and sandstone pressures will rise in equilibrium if the layers have equivalent elastic properties ($\gamma = 1$). Otherwise, the two pressures will increase at different rates, diverging linearly. Figure~\ref{fig:earlytimes} illustrates this early time solution for $\gamma = 2$, meaning that the mudstone is twice as compressible as the sandstone. The reciprocal of compressibility (i.e.,~stiffness) of a rock is a measure of the change in pressure per unit change in strain. Therefore in this case, the sandstone pressurises twice as fast as the mudstone. Mudstones are generally expected to be more compressible than sandstones \citep{chang2013reduction}. However, mudstones are also typically more overpressured than sandstones \citep{osborne1997mechanisms}, contrary to the model result above. This is because typically sandstones leak some overpressure via a baffled pathway; without this leakage, the sandstone would be more overpressured during tectonic compression. In the Levant Basin, the sandstones that feed fluid vents allow overpressure to accumulate until seal failure, indicating that they cannot efficiently leak overpressure. Hence we exclude leakage from our model.

\begin{figure}[!htbp]
\centering
\begin{subfigure}[b]{0.9\textwidth}
   \includegraphics[width=1\linewidth]{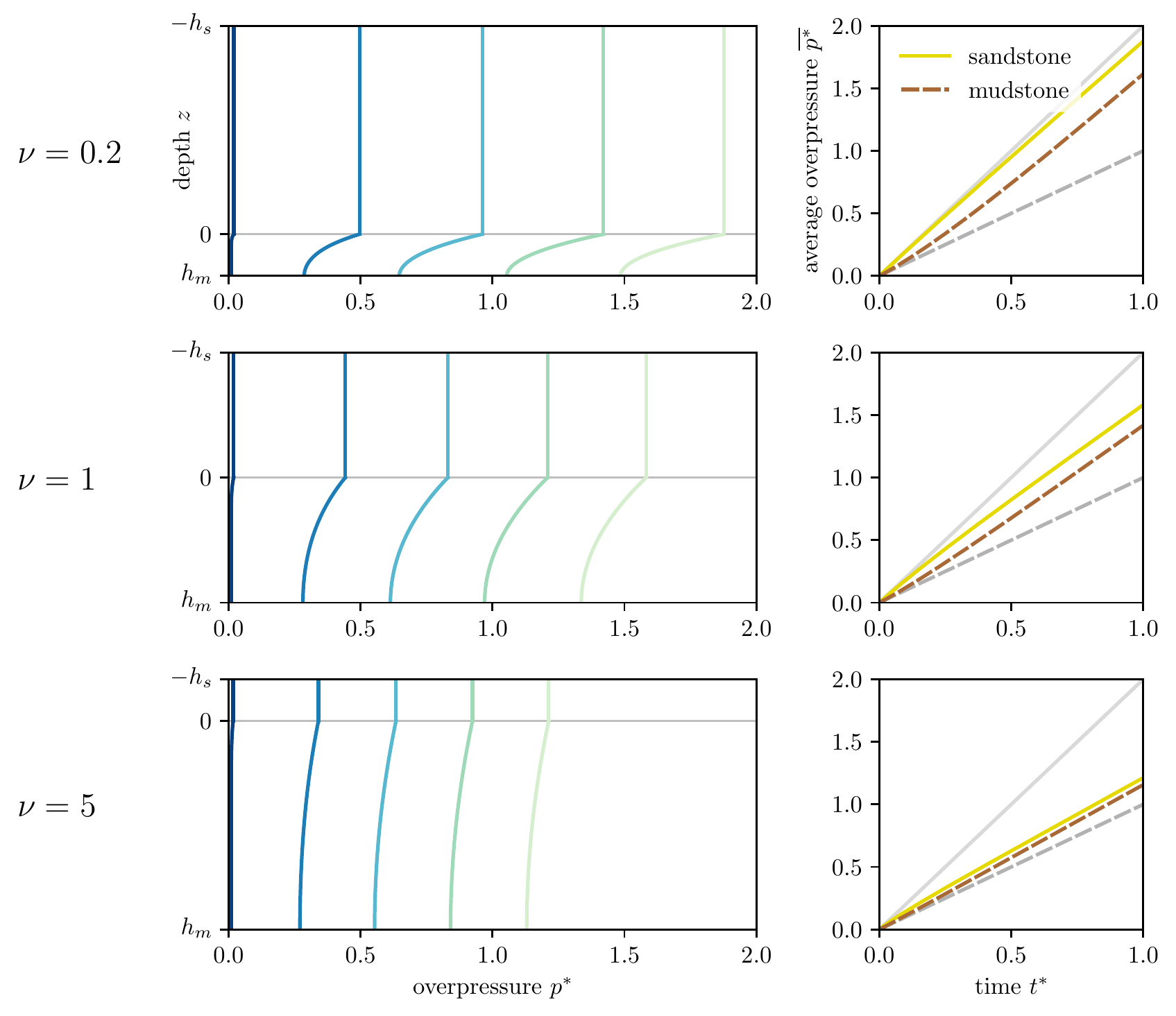}
\end{subfigure}
\caption{Full compression solution for $\gamma = 2$, and three different values of $\nu$. Left: Overpressure versus depth at times $t^* = (0.01, 0.25, 0.5, 0.75, 1)$ (dark to light colours). Right: Time evolution of the average overpressure in the sandstone (solid yellow) and mudstone (dashed brown). We use $\nu$ to indicate the relative thicknesses of each layer, assuming $\nu = h_m/h_s$. We also show the corresponding early-time solutions for each layer (solid grey for sandstone, dashed grey for mudstone).
\label{fig:latetimes}}
\end{figure}

For times $t^* \sim 1$ and larger, the homogenising action of diffusion becomes increasingly important. The system evolves toward a late-time ($t^* \gg 1$) behaviour characterised by
\begin{equation}
    p_m^* \sim \frac{\gamma + \nu}{1 + \nu} t^* + \frac{\nu(\gamma-1)}{3(1 + \nu)^2} - \frac{\gamma-1}{2(1 + \nu)}z^*(2-z^*), \hspace{1cm} p_s^* \sim \frac{\gamma + \nu}{1 + \nu} t^* + \frac{\nu(\gamma-1)}{3(1 + \nu)^2}, \hspace{1cm} \mathrm{for} \,\, t^* \gg 1.
\end{equation}
In this late-time state, the sandstone pressure increases at a rate given by $(\gamma+\nu)/(1+\nu)$, which is between the independent, early-time rates of the two layers. The mudstone pressure tends toward a parabolic profile in $z$. If $\gamma > 1$, the mudstone pressure will be maximum at the sandstone and decrease with depth (Fig.~\ref{fig:latetimes}). If $\gamma < 1$, the mudstone pressure will be minimum at the sandstone and increase with depth.

Fig.~\ref{fig:latetimes} illustrates the impact of $\nu$ on the pressure behaviour at late times, taking $\gamma = 2$ for comparison with the early-time in Fig.~\ref{fig:earlytimes}. Compression pressurises the sandstone more quickly than the mudstone, driving fluid from the sandstone into the mudstone. At $\nu = 5$, the sandstone is small and/or less compressible compared to the mudstone, thus the mudstone pressure will not vary significantly from its compression-driven trajectory while the pressures equilibrate. In the opposite limit ($\nu = 0.2$), this transfer of fluid pressurises the mudstone without noticeably depressurising the sandstone. For any $\nu$, however, both pressures grow in time and will eventually reach the minimum compressive stress (Supplementary Material~\ref{sup:hydfrac}), leading to hydraulic fracturing and fluid venting. We incorporate these phenomena into the model in the next section.

\subsection{Fluid venting}
\label{sec:venting}

\begin{figure}[!htbp]
\centering
\begin{subfigure}[b]{0.99\textwidth}
   \includegraphics[width=1\linewidth]{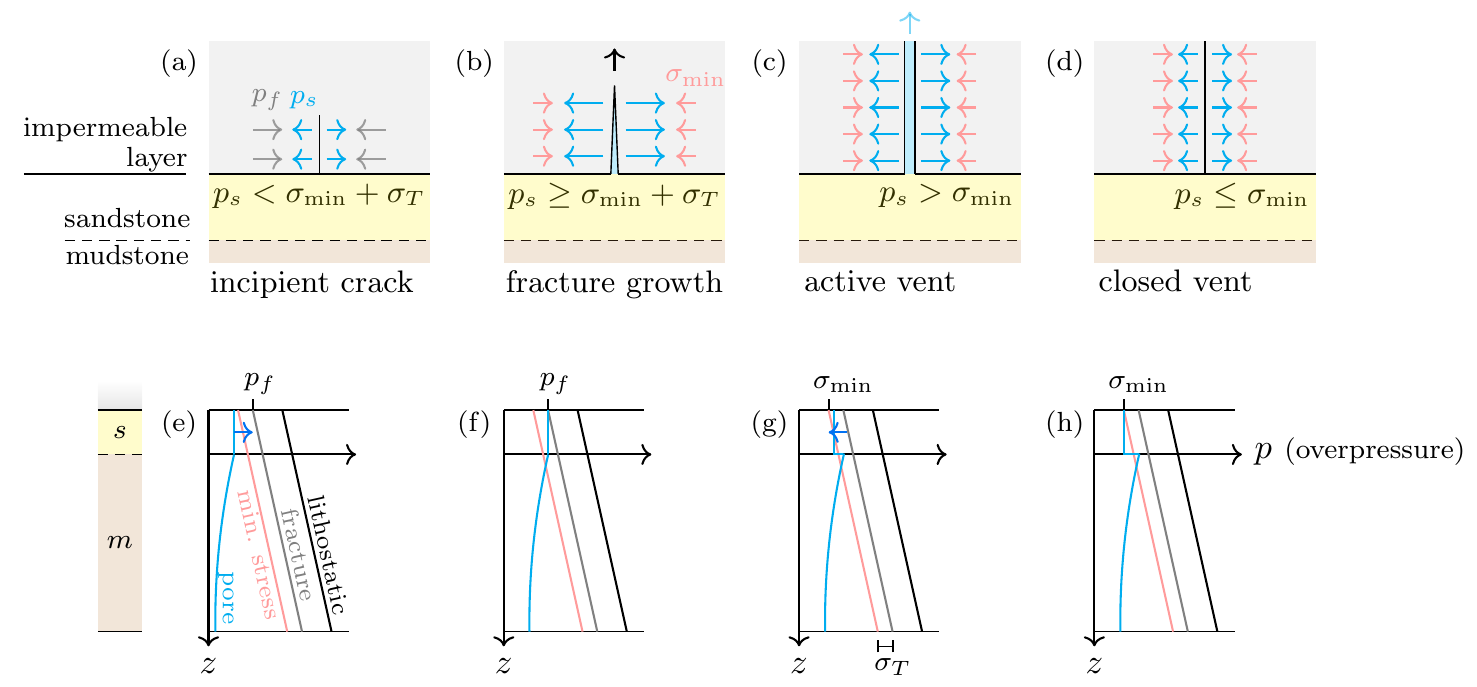}
\end{subfigure}
\caption{
Schematic diagrams of our conceptual model for venting with corresponding overpressure--depth plots. Panels~(a)--(d) are as in Fig.~\ref{fig:model}a, but magnified into the sandstone layer (yellow) and the overlying sealing layer (grey). (a)~A microscopic incipient fracture in the sealing layer will remain closed as long as the sandstone overpressure $p_s$ (blue arrows) is less than the critical fracture overpressure $p_f = \sigma_{\mathrm{min}} + \sigma_T$ (grey arrows), where $\sigma_\mathrm{min}$ here represents the minimum compressive stress minus the hydrostatic pressure. (b)~Once $p_s$ exceeds $p_f$, a fracture will open and grow toward the surface. The fractured rock has no tensile strength, so only $\sigma_\mathrm{min}$ acts against the fracture. (c)~An active vent sources fluid from the sandstone layer and depressurises it until $p_s$ falls back to $\sigma_\mathrm{min}$. (d)~Once $p_s$ equals $\sigma_\mathrm{min}$, the vent closes, allowing the fracture to heal. Panels~(e)--(h) show overpressure--depth plots corresponding to the incipient-crack scenario in (a), the fracture-growth scenario in (b), the active-vent scenario in (c) and the closed-vent scenario in (d), respectively. (e) The pore overpressure in the sandstone $p_s$ is less than the critical fracture overpressure at the top sandstone $p_f$ (grey). (f)~The fracture propagates and opens once $p_s$ exceeds $p_f$. (g)~Venting occurs until $p_s$ falls back to $\sigma_{\mathrm{min}}$. (h)~At which point, the vent closes.
\label{fig:ventmodel}}
\end{figure}

\begin{figure}[!htbp]
\centering
\begin{subfigure}[b]{0.71\textwidth}
   \includegraphics[width=1\linewidth]{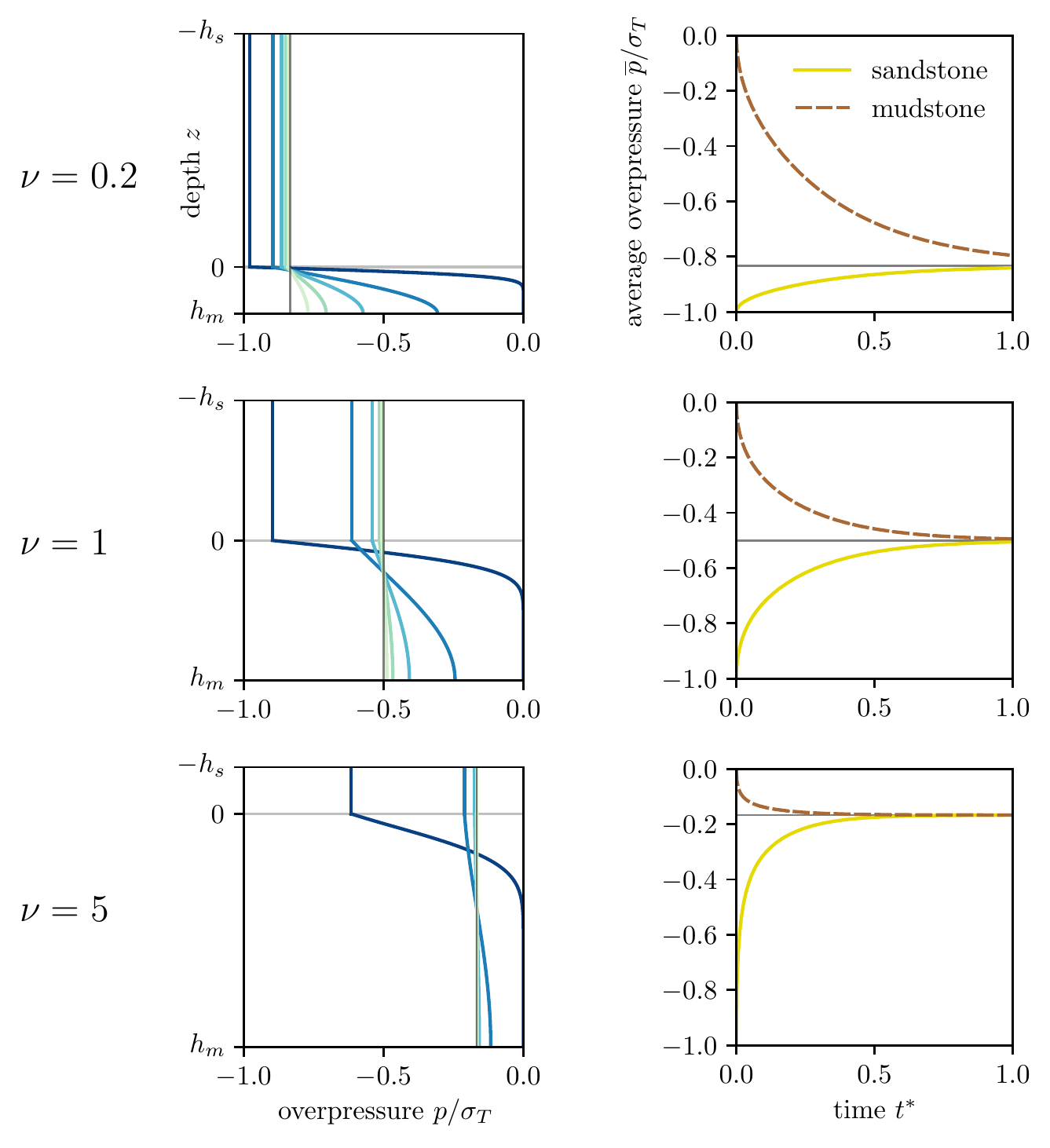}
\end{subfigure}
\caption{Pressure evolution after a venting episode for three different values of $\nu$ in the absence of tectonic compression. Left: Pressure--depth curves at different times with time increasing from dark to light. Right: Average pressure evolution of the sandstone (solid yellow) and mudstone (dashed brown). The late-time equilibrium pressure is also shown (grey).
\label{fig:pipenu}}
\end{figure}

Fluid venting is assumed to occur by hydraulic fracturing. A vertical hydraulic fracture forms when the pore pressure exceeds a critical value $P_f$, given by the sum of the minimum horizontal compressive stress $\sigma_{\mathrm{min}}$ and the tensile strength $\sigma_T$ of the impermeable rock \citep{price1990analysis},
\begin{equation}
    P_f = \sigma_{\mathrm{min}} + \sigma_T.
\end{equation}
Once venting begins, the sandstone pressure drops rapidly until the fracture closes, which we assume occurs when $P_s$ reaches $\sigma_{\mathrm{min}}$, giving an overall pressure drop of $\sigma_T$. Once closed, the fracture heals and may be carried away from the venting point by viscous creep of the salt. Since venting is geologically instantaneous, we incorporate venting into our model by augmenting the (nondimensional) sandstone boundary condition as follows,
\begin{equation}
\label{eqn:pipe_gov2}
    \pd{p_m^*}{t^*} =  -\sigma_T^* \delta(t^* - t_f^*) + \gamma + \nu \pd{p_m^*}{z^*} \hspace{0.5cm} \mathrm{at} \,\, z^*=0.
\end{equation}
where $\sigma_T \equiv \sigma_T^* \, \Gamma_m h_m^2 / D_m$ is a dimensionless tensile strength, $t_f^*$ is the dimensionless time when the sandstone overpressure reaches $p_f^*$ (i.e., $p_m^*(0,t_f^*) \equiv p_f^*$), and $\delta(t^* - t_f^*)$ is the Dirac delta function. This boundary condition imposes an instantaneous drop in sandstone pressure of $\sigma_T$ at time $t_f^*$. The solution to the modified system of equations is separable into two parts: one for compression (i.e., the solution derived above, Eq.~(\ref{eqn:compsol})), and one representing the response to pipe formation. We derive the latter solution for the overpressure response to pipe formation, $\tilde{p}_m$ (dimensional), via Laplace transforms (Supplementary Material~\ref{sup:ventsol}), arriving at
\begin{equation}
\label{eqn:pipesol}
    \frac{\tilde{p}_m}{\sigma_T} = -\frac{1}{1 + \nu} - 2\nu \sum_{j=1}^\infty \frac{\cos \{ \xi_j (1-z^*) \} }{ \cos \xi_j } \frac{\exp\{-\xi_j^2 (t^*-t_f^*)\} }{ {\nu}^2 + \nu + \xi_j^2 }.
\end{equation}
At late times, the rightmost term vanishes as the layer pressures equilibrate by pressure diffusion. In transferring fluid to the depressurised sandstone, the mudstone pressure drops by $\sigma_T/(1+\nu)$, while the sandstone pressure rises by $\sigma_T \, \nu/(1+\nu)$. Increasing $\nu$, for example by increasing the size of the mudstone relative to the sandstone, reduces the sensitivity of the mudstone to venting from the sandstone (Fig.~\ref{fig:pipenu}). An infinitely large mudstone will have a constant mean pressure in response to venting. Increasing $\nu$ also reduces the time needed for pressure equilibration because depressurisation is more localised to the mudstone--sandstone boundary.

Even in the absence of further tectonic compression, pressure diffusion from the mudstone can fully recharge the sandstone layer after venting, potentially multiple times. However, it is likely that compression continues to act once venting has initiated. In the next section, we explore the combined system for compression and venting.

\begin{figure}[!htbp]
\centering
\begin{subfigure}[b]{0.99\textwidth}
   \includegraphics[width=1\linewidth]{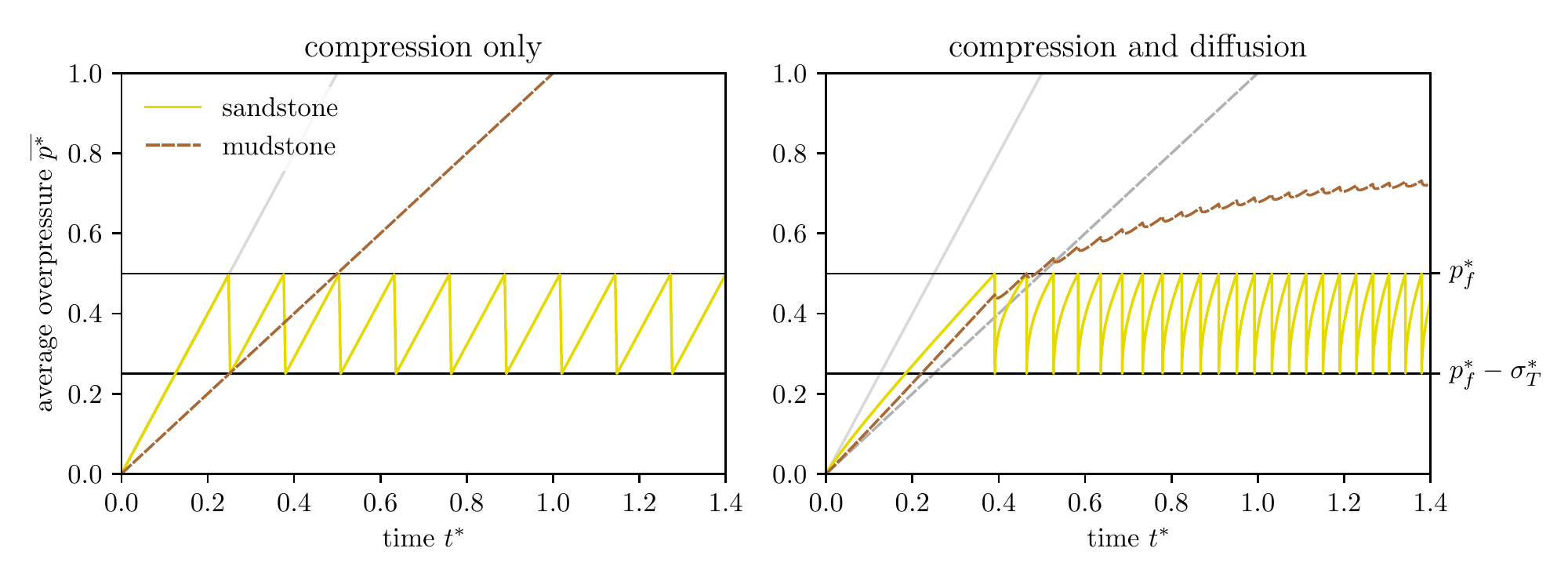}
\end{subfigure}
\caption{Comparison of average overpressure evolution of the sandstone (solid yellow) and mudstone (dashed brown) layers with compression only (left) and with both compression and pressure diffusion (right). Early time solutions without venting are plotted in grey for each layer. Here, $\gamma = 2$, $\nu = 5$, $p_f^* = 0.5$ and $\sigma_T^* = 0.25$. 
\label{fig:diff_vs_no_diff}}
\end{figure}

\subsection{Episodic venting}
\label{sec:episodic}

The solution for simultaneous compression and venting is obtained by superposing the two independent solutions, Eq.~(\ref{eqn:compsol}) for compression and Eq.~(\ref{eqn:pipesol}) for venting. In particular, the linearity of the model allows episodic venting to be represented by a superposition of venting responses at a set of fracture times. The $n^\mathrm{th}$ fracture time $t_{f,n}$ must be determined numerically using the condition $p_m(0,t_{f,n}) = p_f$. An example scenario of continuous compression with venting is illustrated in Fig.~\ref{fig:diff_vs_no_diff} for $\gamma = 2$ and $\nu = 5$, in which case the mudstone is larger and more compressible than the sandstone. 

Without pressure diffusion (i.e., with compression only), our model reduces to the sawtooth model proposed by \cite{cartwright2021quantitative} (Fig.~\ref{fig:diff_vs_no_diff}a). With pressure diffusion, our full model produces qualitatively similar sawtooth behaviour for the sandstone pressure (Fig.~\ref{fig:diff_vs_no_diff}b). However, time between venting episodes (i.e., the venting period) is notably shorter because pressure diffusion from the mudstone provides an additional contribution to pressure recharge. As the mudstone pressure increases relative to the sandstone pressure, this diffusive flux into the sandstone increases and the venting period decreases. The venting period tends towards a constant as the mudstone pressure oscillates around its asymptotic mean value. This transition to periodic behaviour occurs over the characteristic poroelastic timescale $h_m^2/D_m$, i.e.,~over one unit of dimensionless time (Supplementary Material \ref{sec:transition}). For example, Figure~\ref{fig:diff_vs_no_diff}b shows that after the onset of venting at $t^* \approx 0.4$, the mudstone pressure effectively reaches its asymptotic mean value at $t^* \approx 1.4$. If the poroelastic timescale is much greater than the duration of venting in the system, then pressure diffusion is negligible and the sawtooth model is a good approximation. If instead the poroelastic timescale is much shorter than the venting period, periodic venting is established immediately after the first episode. In this periodic-venting limit, we can recover explicit solutions for behaviour of this system.

\subsubsection{Periodic venting}

\begin{figure}[!t]
\centering
\begin{subfigure}[t]{0.75\textwidth}
   \includegraphics[width=1\linewidth]{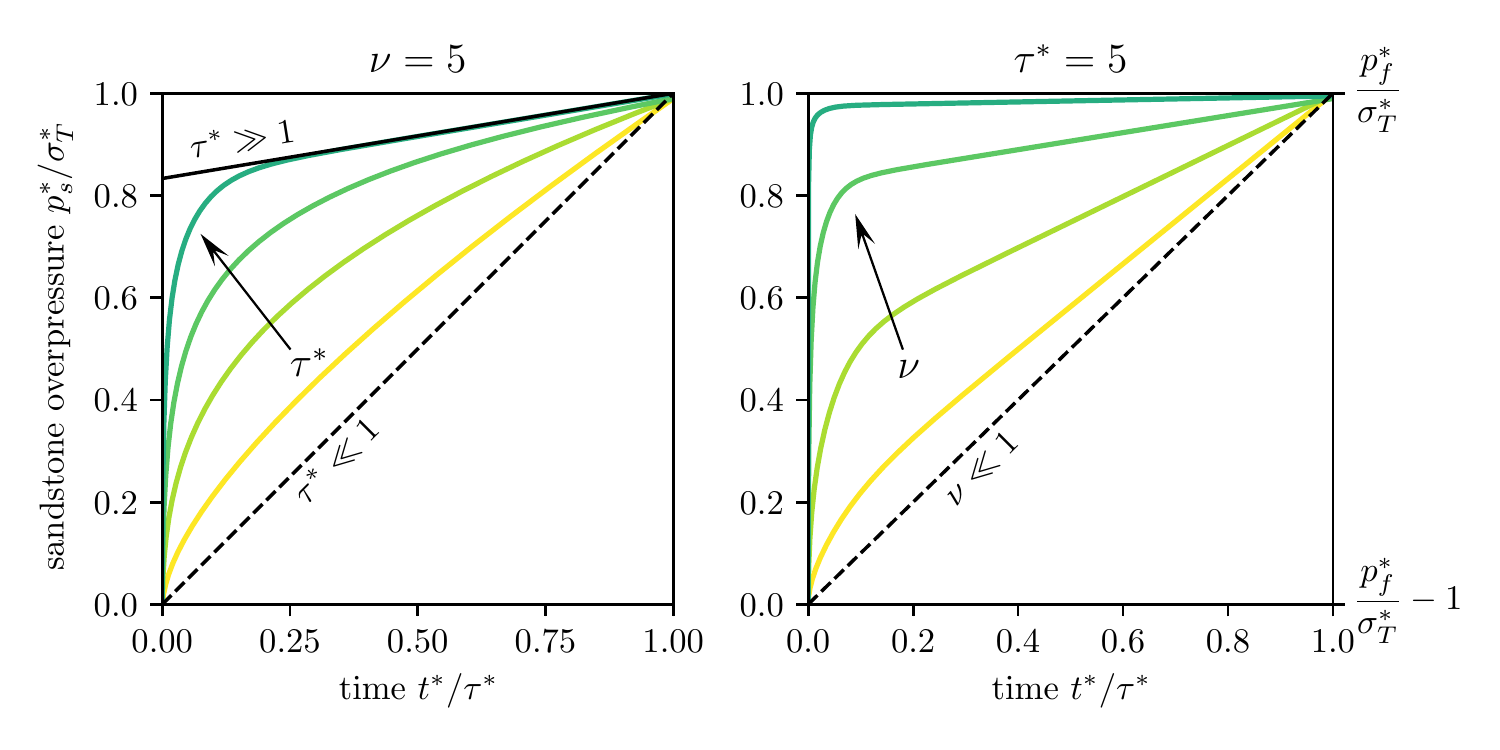}
\end{subfigure}
\caption{Sandstone pressure recharge behaviour during periodic venting, with varying $\tau^*$ and $\nu$. We normalise dimensionless time $t^*$ by the dimensionless venting period $\tau^*$, such that $t^*/\tau^* \in [0,1]$ over one period. Similarly, the dimensionless sandstone pressure $p_s^*$ is scaled by the dimensionless tensile strength $\sigma_T^*$. With this normalisation, Eq.~(\ref{eqn:perpress}) is explicitly controlled by two dimensionless parameters, $\nu$ and $\tau^*$. (a) Varying $\tau^* = (0.01, 0.1, 0.5, 2.5)$ at $\nu = 5$ with limiting behaviours for $\tau^* \ll 1$ (dashed black) and $\tau^* \gg 1$ (solid black). (b) Varying $\nu = (0.2, 1, 5, 50)$ at $\tau^* = 5$ with an limiting behaviour for $\nu \ll 1$ (dashed).
\label{fig:persol}}
\end{figure}

Using the principle of superposition, the periodic solution for pressure in the sandstone is
\begin{align}
\label{eqn:perpress}
    p_s^* = \frac{\gamma + \nu}{1 + \nu} t^* + 2\nu \sigma_T^*\sum_{n=0}^\infty\sum_{j=1}^\infty \frac{1 - \exp(-\xi_j^2 t^*) }{{\nu}^2 + \nu + \xi_j^2 }\exp(-\xi_j^2 n \tau^*),
\end{align}
where $\tau^*$ is the dimensionless venting period, scaled with the characteristic time for pressure diffusion $h_m^2/D_m$ (and hence the dimensional venting period is $\tau \equiv \tau^* \, h_m^2/D_m$). The periodic venting solution~\eqref{eqn:perpress} is a function of two dimensionless parameters, $\nu$ and $\tau^*$. Figure~\ref{fig:persol} shows how these parameters control the periodic recharge pathway. 

Solving Eq.~\eqref{eqn:perpress} for $\tau^*$ (Supplementary Material~\ref{sup:period}) gives
\begin{equation}
\label{eqn:taunondim}
    \tau^* = \frac{\sigma_T^*}{\gamma + \nu}.
\end{equation}
If $\tau^* \ll 1$ (i.e., if the dimensional venting period $\tau$ is much shorter than the diffusive time $h_m^2/D_m$), tectonic compression is the dominant contribution to pressure recharge and we recover a sawtooth; this is shown by the dashed black curve in Fig.~\ref{fig:persol}a. If instead the venting period is much longer than the diffusive time, $\tau^* \gg 1$, diffusion equilibrates the sandstone and mudstone pressures to $p_s/\sigma_T = p_m/\sigma_T = \nu/(1+\nu)$; both then rise at the same rate, due to tectonic compression (solid black curve in Fig.~\ref{fig:persol}a).

\begin{figure}[!htbp]
\centering
\begin{subfigure}[b]{0.7\textwidth}
   \includegraphics[width=1\linewidth]{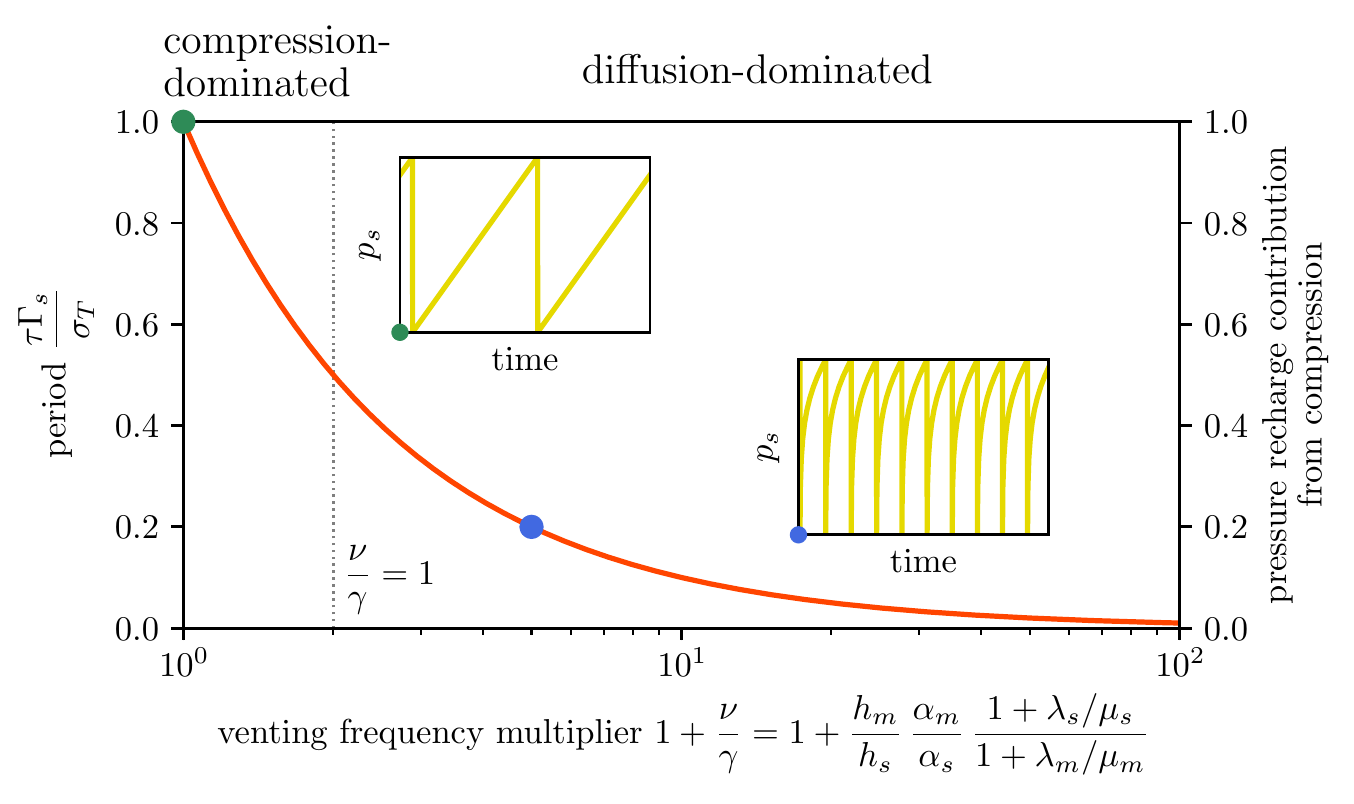}
\end{subfigure}
\caption{Venting period (left axis) and fraction of total pressure recharge from compression (right axis) as functions of the venting frequency multiplier, $(1 + \nu/\gamma)$, where the dimensionless ratio $\nu/\gamma$ measures the pressure recharge contribution of pressure diffusion relative to that of compression. The two quantities have the same mathematical relationship with $(1 + \nu/\gamma)$ and therefore follow the same curve. The same function is the pressure recharge contribution from tectonic compression. Insets show example compression-dominated (green dot) and diffusion-dominated (blue dot) pressure evolution curves for the sandstone.
\label{fig:nugamma}}
\end{figure}

These diffusive effects are also controlled by the capacitance ratio $\nu$ of the two layers. If the hydraulic capacitance of the mudstone is negligible relative to that of the sandstone ($\nu \ll 1$), we again recover the sawtooth model (dashed black curve in Fig.~\ref{fig:persol}b). Increasing $\nu$ (i.e.,~increasing the mudstone size) leads to faster diffusive pressure equilibration and a higher equilibration pressure, much like in Fig.~\ref{fig:pipenu}. However, increasing $\nu$ also decreases the compressive pressurisation rate of the equilibrated sandstone and mudstone layers, as shown in Fig.~\ref{fig:latetimes}. This means that a family of compressive and diffusive pressure recharge rates can produce the same venting period. Consequently, observational measurement of the mean venting period alone is insufficient to determine the dominant recharge mechanism. However, the dominant recharge mechanism can be inferred using the expression for venting period in combination with estimates of basin properties. To this end, we rewrite Eq.~(\ref{eqn:taunondim}) in dimensional terms as
\begin{equation}
\label{eqn:taudim}
    \tau = \frac{\sigma_T/\Gamma_s}{1 + \nu/\gamma},
\end{equation}
showing that the period is determined by the tensile strength of the impermeable seal, the tectonic pressurisation rate in the sandstone and $(1+\nu/\gamma)$. This denominator is termed the venting frequency multiplier, where $\nu/\gamma$ is a dimensionless number that quantifies the effect of diffusion. Without pressure diffusion (i.e., for $\nu/\gamma \ll 1$), the frequency multiplier is unity and the period reduces to the compression-only (sawtooth) period, $\sigma_T/\Gamma_s$ (Fig.~\ref{fig:nugamma}). Considering the role of pressure diffusion divides the compression-only period or, equivalently, multiplies the compression-only frequency by a factor of (1 + $\nu/\gamma$). If the frequency multiplier is much greater than unity, neglecting its effect will lead to an erroneous diagnosis of the tectonic compression rate from an observed venting period.

The results above enable us to estimate the contribution to pressure recharge from compression relative to that of pressure diffusion over a single period $\tau$. In the absence of diffusion, compression will generate a pressure of $\gamma \tau^*$ per period.  By definition, the overpressure at the end of one period is the amount needed to trigger venting, $\sigma_T^*$, so the relative contributions from compression and diffusion are $\gamma \tau^* / \sigma_T^* = (1+\nu/\gamma)^{-1}$, and $(\sigma_T^* - \gamma \tau^*)/\sigma_T^* = \nu/\gamma \, (1+\nu/\gamma)^{-1}$, respectively, having used Eq.~(\ref{eqn:taunondim}). The ratio of the contribution from diffusion to the contribution from compression is then
\begin{equation}
\label{eqn:nugamma}
    \frac{\nu}{\gamma} = \frac{h_m}{h_s} \frac{\alpha_m}{\alpha_s} \frac{1 + \lambda_s/\mu_s}{1 + \lambda_m/\mu_m}.
\end{equation}
This dimensionless ratio is the ratio of fluid volumes expelled from each layer during tectonic compression at constant pore pressure. Remarkably, this quantity is independent of strain rate and permeability; it depends only on the thicknesses and poroelastic properties of the layers. Permeability plays an important role in the timescales associated with compression and diffusion, and thus in the shape of the recharge curve, but does not affect the overall contributions. The dimensionless quantity $(1 + \lambda/\mu)$ can be rewritten as $1/(1-2v)$ where $v$ is the Poisson ratio. We can simplify Eq.~(\ref{eqn:nugamma}) by recognising that in most mudstone--sandstone successions, $v_s \sim v_m$, implying that $(1+\lambda_m/\mu_m)/(1+\lambda_s/\mu_s) \sim O(1)$, and the ratio of Biot coefficients $\alpha_m/\alpha_s \sim O(1)$ so that
\begin{equation}
    \frac{\nu}{\gamma} \sim \frac{h_m}{h_s}.
\end{equation}
In other words, the role of pressure diffusion relative to compression is controlled primarily by the thickness of the mudstone relative to that of the sandstone. This equation means that for a mudstone unit that is thick compared to the adjacent sandstone, diffusion dominates the pressure recharge. Fluid venting phenomena are frequently observed in mudstone-dominated basins \citep{cartwright2015seismic}, so pressure diffusion is likely to be a major contributor to pressure recharge for fluid venting phenomena.

\section{Discussion}
\label{sec:discussion}

The expression derived above for the venting period can be used to interpret observations of episodic venting and to infer the history of tectonic compression. As a prototype of such analysis, we consider the Oceanus pockmark trail located in the North Levant Basin \citep{cartwright2018direct}. \cite{cartwright2021quantitative} performed a similar analysis, beginning with the assumption that the associated reservoir was hydrostatically pressured (i.e.,~zero overpressure) immediately after the Messinian Salinity Crisis. They then estimated that $\sim$30~MPa of overpressure must have been generated over a period of $\sim$3~Myr to exceed the critical fracture pressure and initiate the first venting episode, implying a pressurisation rate of $\sim$10~MPa/Myr prior to initiation of venting. They attributed this pressurisation rate solely to tectonic compression.

The Oceanus trail records 21 venting episodes from $\sim$1.7~Ma until recent, suggesting a mean venting period of $\sim$80~kyr. During each venting episode, the Messinian salt layer is re-fractured and fluid is expelled until the sandstone pressure drops by the tensile strength of the salt, as described above. Extended leak-off tests suggest that this tensile strength is in the range 2~$\pm$~1~MPa \citep{berest2015maximum}. Consequently, a pressurisation rate of 25~$\pm$~13~MPa/Myr would be required to recharge the reservoir to failure every 80~kyr. At Oceanus, the pre-salt clastic succession from which the fluid is sourced is predominantly mudstone, with $h_m/h_s$~$\sim$~20~$\pm$~7 \citep{cartwright2021quantitative}. Using typical sandstone and mudstone properties from Table~\ref{tab:nugamma} in Eq.~(\ref{eqn:nugamma}), we estimate the venting frequency multiplier to be $(1+\nu/\gamma)\sim$~24$~\pm$~12. With this result, we use Eq.~(\ref{eqn:taudim}) to calculate the pressurisation rate due to tectonic compression during the periodic-venting phase to be $\Gamma_s \sim$~1.4~$\pm$~0.9~MPa/Myr. This result is much lower than the total inferred pressurisation rate during periodic venting because the major recharge contribution is from pressure diffusion, with a rate of 24~$\pm$~14~MPa/Myr. Additionally, the inferred pressurisation rate due to tectonic compression during periodic venting is much lower than that obtained by \cite{cartwright2021quantitative} for the period prior to the onset of venting. One possible explanation for this discrepancy is that tectonic compression has eased since the initiation of venting. However, it is also plausible that the Messinian Salinity Crisis left the Oceanus reservoir significantly overpressured \citep{al2016impact}, rather than hydrostatically pressured, and that the rate of tectonic compression has been roughly constant since that time.

\begin{table}[]
\centering
\begin{tabular}{lllllll}
\textbf{Parameter}              & \textbf{Description}       & \textbf{mean} & \textbf{std. dev.} & \textbf{min.} & \textbf{max.} & \textbf{Reference} \\ \hline
$\sigma_T$ (MPa)                       & salt tensile strength        & 2           & 1                 & 0.5            & 4           & \cite{berest2015maximum} \\ \hline
$h_s$ (m)                       & sandstone thickness        & 150           & 50                 & 50            & 200           & \cite{cartwright2021quantitative}, \\
$h_m$ (m)                       & mudstone thickness         & 2500          & 250                & 2000          & 3000          &  Fig. 2b                              \\ \hline
$\alpha_s$                      & sandstone Biot coefficient & 0.62          & 0.17               & 0.38          & 0.83          & \cite{ge1992hydromechanical}                              \\
$\alpha_m$                                                                                                                               & mudstone Biot coefficient  & 0.68          & 0.35               & 0.30          & 0.98          &                               \\
$v_s$           & sandstone Poisson ratio    & 0.24          & 0.04               & 0.20          & 0.30          &                               \\
$v_m$           & mudstone Poisson ratio     & 0.25          & 0.05               & 0.15          & 0.30          &                               \\
$K_s$ (GPa)                     & sandstone bulk modulus     & 18            & 8                  & 8             & 30            &                               \\
$K_m$ (GPa)                     & mudstone bulk modulus      & 15            & 17                 & 5             & 33            &                               \\ \hline
$\phi_s$                        & sandstone porosity         & 0.22          & 0.01               & 0.19          & 0.24          & \cite{ortega2018dynamic}                              \\ \hline
$\phi_m$                        & mudstone porosity         & 0.20          & 0.05               & 0.05          & 0.30          &  \cite{yang2007permeability} \\
log$_{10}$$k_m$ (log$_{10}\,$m$^2$)                       & (log) mudstone permeability         & -19          & 0.5                & -22          & -18          &                               \\ \hline
$\eta$ (mPa$\,$s)                       & water viscosity         & 0.3          & 0.1              & 0.1          & 0.5          & \cite{abramson2007viscosity} \\ \hline
$c_\ell$ (10$^{-11}$ Pa$^{-1}$) & water compressibility      & 4.0           & 0.1                & 3.7           & 4.3           & \cite{fine1973compressibility} \\[2mm]
\end{tabular}
\caption{Sandstone and mudstone properties used in estimating the compression rate at the Oceanus pockmark trail in the North Levant Basin. We assign a truncated normal distribution for each parameter, taking all parameters to be uncorrelated. The mean and standard deviation of each distribution, as well as the minimum and maximum values at which the distributions are truncated, are given by the values presented in the table. From these distributions, we use the model under a Monte Carlo framework to calculate a distribution for the local strain rate at Oceanus.
\label{tab:nugamma}}
\end{table}

Tectonic compression of the North Levant Basin stems from activity on the Dead Sea Transform fault, where geodetic strain rates of order $10^{-16}$~s$^{-1}$ are observed \citep{palano2013current}. From the pressurisation rate calculated above, we can infer the strain rate associated with tectonic compression at the Oceanus pockmark trail. Using typical sandstone properties (Table~\ref{tab:nugamma}), we use Eq.~(\ref{eqn:gammadef}) to estimate a horizontal strain rate of 1.6~$\pm$~1.1~$\times$~10$^{-4}$~Myr$^{-1}$ = 5.1~$\pm$~3.7~$\times$~10$^{-18}$~s$^{-1}$. This value is much smaller than the observed rate of regional shortening because folding and thrusting across the basin \citep{oppo2021leaky} accommodate a large portion of the shortening. Only the much smaller fraction accommodated by compression is relevant to our model. Strain rates typically vary laterally across sedimentary basins, so this measurement only provides a local strain rate at the Oceanus pockmark trail. 

The above analysis of the Oceanus pockmark trail relies upon the assumption that the transition to periodic venting is complete. This transition occurs after the onset of venting over the characteristic poroelastic timescale of the system $h_m^2/D_m$ (Supplementary Material \ref{sec:transition}). For the Levant Basin, using estimated values from Table~\ref{tab:nugamma}, we calculate an expected poroelastic timescale of 23~kyr. Within the 95\% confidence interval, however, the poroelastic timescale could be an order of magnitude higher or lower. This range is primarily due to the uncertainty in the mudstone permeability. For the typical $\sim$100~kyr venting period, this implies that the transition to periodic venting is complete by the second venting episode or, at most, after a few episodes. Venting observations from the Oceanus trail and other trails across the North Levant Basin show no evidence of an initial transition to periodic venting. This suggests that the regional poroelastic timescale is similar to or less than the $\sim$100~kyr venting period and that our assumption of periodicity is justified.

\cite{oppo2021leaky} documented a set of 12 pockmark trails across the margin of the North Levant Basin, where venting initiates at a similar time to Oceanus. Mean venting periods vary between trails, ranging from $\sim$30 to $\sim$220~kyr. Equation~(\ref{eqn:taudim}) suggests that these spatial variations in period can most likely be attributed to spatial variations in layer thicknesses and the local tectonic strain rate. Moreover, the time of the first venting episode varies significantly between trails; this is controlled by the local tectonic strain rate and the residual overpressure from the Messinian Salinity Crisis. The time of the last observed venting event also varies between trails, potentially associated with the time that compression ceases locally. However, a given reservoir may continue to vent after tectonic compression ceases, driven by pressure diffusion alone (Supplementary Material~\ref{sup:residvent}). The number of excess venting episodes depends on the local hydraulic capacitance ratio and salt tensile strength.

\cite{oppo2021leaky} showed that pipe trails have formed exclusively above anticlines across the North Levant Basin. From this, one might infer that folding generates sufficient overpressure for venting, despite our assumption of the contrary for the present model. \cite{cartwright2021quantitative} estimate that growth of the Oceanus anticline has increased the reservoir trap capacity by 320~m since the initiation of venting. Anticline growth generates overpressure in a high-permeability reservoir through the lateral transfer of pressure to its crest from the surrounding overpressured, low-permeability rock, which is the mudstone in this case \citep{flemings2002flow}. If the reservoir has a parabolic fold geometry with constant thickness, then the corresponding increase in overpressure is $\tfrac{1}{3} (\rho_m - \rho_g) g \Delta h$, where $\rho_m$ is the density of the mudstone, $\rho_g$ is the density of gas, $g$ is the acceleration due to gravity, and $\Delta h$ is the increase in trap capacity. Using density values from \cite{cartwright2021quantitative}, we obtain an increase in overpressure by lateral transfer of $\sim$2~MPa. Additionally, an increase in trap capacity will enable the accumulation of a thicker gas column, increasing overpressure due to buoyancy by $(\rho_g - \rho_w) g \Delta h \sim 1$~MPa, where $\rho_w$ is the density of water. The overpressure generated in the sandstone due to fold growth is thus negligible, merely supplying enough overpressure for one additional venting episode. Despite this, the additional overpressure (and high topography, hence reduced compressive stress) associated with anticlines make them preferential sites for venting.

However, the overpressure generated by fold growth is comparable to that generated by tectonic compression, inferred to be 2.4~$\pm$~1.5~MPa since the initiation of venting. The fundamental difference between these overpressure mechanisms is that tectonic compression also pressurises the mudstone. For example, an overpressure increase in the mudstone of 1~MPa can provide an increase in overpressure in the sandstone of $\nu$~MPa via pressure diffusion. In mudstone-dominated basins $\nu \gg 1$, thus a small amount of mudstone overpressure can drive many venting episodes. This suggests that overpressure mechanisms that pressurise the mudstone, e.g.,~tectonic compression, disequilibrium compaction and sea-level change are potent drivers of episodic fluid venting in other sedimentary basins.

Our model can speculatively be applied to basins containing mud volcanoes, where a sandstone layer may not be present. To do so, the modelled sandstone layer can be recast as a fluidised mud region that depressurises during venting. The thickness of the depressurised region depends on details of the fracture mechanics of venting, and may vary as mud is expelled. If this region is much smaller than the overall mudstone thickness, then it is likely that pressure diffusion provides the dominant contribution to recharge. In the Eastern Mediterranean, the West Nile Delta hosts linear trails composed of fluid escape pipes and mud volcanoes. Sandstone reservoirs are not observed, in contrast to the North Levant Basin. The West Nile Delta is likely pressurised by hydrocarbon generation and lateral pressure transfer rather than tectonic compression \citep{kirkham2022episodic}. Despite these differences, similar $\sim$100~kyr periods are observed in the West Nile Delta \citep{kirkham2022episodic}. This timescale is comparable to the poroelastic diffusion time for the North Levant Basin, suggesting that recharge in this region is also controlled by pressure diffusion.

The model is readily generalisable to other settings and scenarios. For example, the impermeable sealing layer could be replaced with a permeable unit (e.g.,~mudstone) to study the effect of pressure dissipation due to fluid migration through an imperfect seal. The model could also be reformulated to include disequilibrium compaction as the active overpressure mechanism (Supplementary Material~\ref{sup:diseq}). Disequilibrium compaction is a prevalent driver of overpressure in low-permeability sediments \citep{osborne1997mechanisms}, especially in basins with high sedimentation rates \citep[e.g.,~the South Caspian Basin, see][]{stewart2006structure}. For this mechanism, an increasing vertical stress acts on the system, rather than a horizontal compression. Much of our qualitative findings will also apply for disequilibrium compaction. 

Despite the widespread occurrence of venting phenomena, there are many basins comprising overpressured mudstones that have not vented. Venting requires a mechanism that generates overpressure faster than fluid flow can dissipate it, and that is active for long enough to achieve critical overpressures. Therefore, lateral reservoir connectivity and/or poor vertical sealing can inhibit fluid venting. Venting is also less likely to initiate at larger depths, where the critical fracture pressure is larger. Once venting initiates, further episodes require pressure recharge in the reservoir unit, from either sustained overpressure generation or by pressure diffusion from an adjacent overpressured mudstone.

Our model invokes a variety of simplifying assumptions. These assumptions make the model tractable and transparent, but also carry limitations. In reality, tectonic compression is unlikely to be a constant-strain-rate process and compression may not be the sole overpressure mechanism in a basin. Furthermore, basin stratigraphies seldom consist of uniform, horizontal layers; layer topography and heterogeneity will affect the spatial distribution of overpressure and fracture pressure, and thus also of venting loci. The sedimentary layers themselves are composed of rocks with porosity-dependent permeability that do not behave perfectly poro-elastically. For example, many sedimentary rocks may partially fluidise during sudden fluid expulsion. We have assumed that the vented fluid is water, but in reality it is a mixture of water, natural gas and sediment. However, we do not expect any of these features to change the key qualitative implications of our model.

Many open questions remain regarding fluid venting phenomena that are not addressed by this work. The mechanics of venting are not explicitly stated in the model; the high permeability of the sandstone layer enables us to circumvent this. If the sandstone is also overlain by mudstone, then we might expect that hydraulic fracturing occurs through the mudstone. But the mudstone is also predicted to be overpressured and to recharge the reservoir during episodic venting. It is unclear how a hydraulic fracture could propagate through a unit with a higher overpressure; perhaps the presence of natural gas and fluidised sediment enables this. This lack of clarity in the conditions for venting motivates incorporation of fracture mechanics and mudstone fluidisation into our model. Furthermore, since the model contains only one spatial dimension, it cannot be used to address observations such as the strong correlation of venting sites with areas of high reservoir-layer topography. This strong spatial dependence can lead to clusters of multiple venting sites in close proximity. For example, in the North Levant Basin margin \citep{oppo2021leaky}, four pockmark trails originate along the crest of the same folded sandstone reservoir and thus may be hydraulically connected. A model that considers these spatial variations may be capable of inferring basin properties such as reservoir architecture from dates of venting episodes.

\section{Conclusions}
\label{sec:conclusions}

We have developed and solved a one-dimensional model of pressure evolution of sedimentary layers subject to horizontal tectonic compression, leading to episodic fluid venting. Our main goal was to elucidate the interactions between the three fundamental ingredients of episodic venting: pressure build-up, pressure diffusion and hydraulic fracturing. The main conclusions from the study are as follows.

\begin{itemize}
  \item In the absence of hydraulic fracturing, two time regimes emerge. At early times, the pressure of each layer rises independently at a rate determined by the elastic properties of the layer. At timescales corresponding to mudstone fluid flow ($t \gg h_m^2/D_m$), layer pressures equilibrate by pressure diffusion and rise at the same rate.
  
  \item Sustained tectonic compression will generate extreme overpressures required for hydraulic fracturing of the sealing layer. Once hydraulic fracturing initiates, the sandstone pressure drops rapidly. The sandstone pressure is then slowly recharged by tectonic compression and pressure diffusion from the mudstone layer, leading to repeated venting episodes.
  
  \item During episodic venting, the time interval between episodes tends towards a fixed period, given by $\tau = \sigma_T/\Gamma_s/(1 + \nu/\gamma)$, where $\sigma_T$ is the tensile strength of the seal, $\Gamma_s$ is the pressure build-up rate in the sandstone layer due to tectonic compression, and $(1 + \nu/\gamma)$ is the venting frequency multiplier. 
  
  \item The venting frequency multiplier $(1 + \nu/\gamma)$ is determined by the dimensionless ratio $\nu/\gamma$, which is independent of strain rate and permeability. When $\nu/\gamma$ is small, diffusion is negligible compared to compression and the sandstone pressure evolves as a sawtooth wave with a venting period $\tau \sim \sigma_T/\Gamma_s$. The period decreases with increasing $\nu/\gamma$ as the additional contribution to pressure recharge from diffusion increases. Based on estimates of sandstone and mudstone elastic properties, $\nu/\gamma$ is controlled by the ratio of thicknesses of the mudstone and sandstone layers. Fluid venting phenomena are commonly found in mudstone-dominated basins; we have shown that in these settings, pressure recharge will be dominated by diffusion. 
  
  \item More generally, we have shown that tectonic compression cannot be decoupled from pressure diffusion. Episodic expulsion depends on more than simply the rate of tectonic compression --- pressure diffusion can markedly reduce the time interval between episodes.
\end{itemize} 
%

\bibliographystyle{agsm}
\renewcommand\bibname{\large{References}}
\bibliography{Sections/references.bib}

\newpage

\renewcommand{\appendixname}{Supplementary Information}
\setcounter{section}{0}
\renewcommand{\thesection}{S\arabic{section}}   

\section*{\titlefont Supplementary material}
\vspace{5mm}
\renewcommand{\thefigure}{S\arabic{figure}}
\setcounter{figure}{0}
\renewcommand{\thetable}{S\arabic{table}}
\setcounter{table}{0}

\section{Storage equation}
\label{sup:storage}

Here, we follow the derivation of the storage equation by \cite{verruijt1969elastic}. Starting from conservation of mass of the fluid $\ell$ and solid $g$ phases,
\begin{equation}
    \pd{\phi \rho_\ell}{t} + \Div \phi \rho_\ell \vel_\ell = 0,
\end{equation}
\begin{equation}
    \pd{(1-\phi) \rho_g}{t} + \Div (1-\phi) \rho_g \vel_g = 0,
\end{equation}
where $\phi$ is the porosity of the medium, $\rho_i$ is the phase density and $\vel_i$ is the phase velocity. Assuming spatial variations in phase densities are negligible, then
\begin{equation}
\label{eqn:boussell}
    \frac{1}{\rho_\ell}\pd{\phi \rho_\ell}{t} + \Div \phi \vel_\ell = 0,
\end{equation}
\begin{equation}
\label{eqn:boussg}
    \frac{1}{\rho_g}\pd{(1-\phi) \rho_g}{t} + \Div (1-\phi) \vel_g = 0.
\end{equation}
For the solid phase, $c_g$ is the compressibility of the solid grains such that for the pure solid material, $c_g \equiv 1/\rho_g \, \partial\rho_g/\partial p$. However, in poroelastic materials, grains compress in response to changes in both total isotropic stress $\sigma$ and pore pressure $p$. 

We determine this relationship by considering an undrained compression of $\Delta p$ followed by a drained compression of $(-\Delta \sigma) - \Delta p$ such that the overall change in pore pressure is $\Delta p$ and the overall change in isotropic stress is $-\Delta \sigma$ (stress is tension-positive). For an undrained deformation of magnitude $\Delta p$, the change in grain density $\Delta \rho_g = c_g \rho_g \Delta p$, by definition. For a drained compression of magnitude $(-\Delta \sigma) - \Delta p$, the grains experience an amplified stress increase of $-(\Delta p + \Delta \sigma)/(1-\phi)$, so
\begin{equation}
    \Delta \rho_g = -c_g \rho_g \frac{\Delta p + \Delta \sigma}{1 - \phi}.
\end{equation}
Superposition of these two responses gives the change in grain density,
\begin{equation}
    \Delta \rho_g = -\frac{c_g\rho_g}{1 - \phi}(\phi \Delta p + \Delta \sigma).
\end{equation}
If these changes occur over a time $\Delta t$, the limit as $\Delta t \rightarrow 0$ is
\begin{equation}
\label{eqn:rhog}
    \pd{\rho_g}{t} = -\frac{c_g\rho_g}{1 - \phi}\bigg(\phi \pd{p}{t} + \pd{\sigma}{t}\bigg).
\end{equation}
Using (\ref{eqn:rhog}), we can simplify the solid mass conservation equation (\ref{eqn:boussg}) to
\begin{equation}
\label{eqn:phig}
    -\pd{\phi}{t} - \phi c_g \pd{p}{t} - c_g \pd{\sigma}{t} + \Div (1-\phi) \vel_s = 0.
\end{equation}
Applying the definition of fluid compressibility $c_\ell \equiv 1/\rho_\ell \, \partial \rho_\ell / \partial p$ to (\ref{eqn:boussell}) gives
\begin{equation}
\label{eqn:phiell}
    \pd{\phi}{t} + \phi c_\ell \pd{p}{t} + \Div \phi \vel_\ell = 0.
\end{equation}
We add (\ref{eqn:phig}) and (\ref{eqn:phiell}) to eliminate the porosity derivative terms, resulting in
\begin{equation}
     \pd{e}{t} + \phi (c_\ell - c_g) \pd{p}{t} -c_g\pd{\sigma}{t} = -\Div \segflux,
\end{equation}
where $\segflux \equiv \phi(\vel_\ell - \vel_g)$ and $\partial e/\partial t \equiv \Div \vel_g$. We can eliminate the isotropic total stress from the equation using Terzaghi's principle, $\sigma = \sigma' - \alpha p$ where the effective stress $\sigma' = e/c$ and $\alpha \equiv 1 - c_g/c$ with $c$ denoting the compressibility of the solid skeleton. It follows that
\begin{equation}
\label{eqn:storage_sm}
      \alpha\pd{e}{t} + S \pd{p}{t} = -\Div \segflux,
\end{equation}
where the storativity $S \equiv \phi c_\ell + (\alpha - \phi) c_g$. Equation (\ref{eqn:storage_sm}) is termed the storage equation by \cite{verruijt1969elastic}. The only assumptions made in deriving this equation are that spatial variations in phase densities are negligible and that each phase is linearly compressible.

\section{Stress conditions for hydraulic fracturing}
\label{sup:hydfrac}

For this analysis, we assume that the initial compressive stress in the $y$-direction is less than the lithostatic stress. As compression is acting in the $x$-direction, a vertical fracture will open aligned with the $y$-direction. In this case, the condition for hydraulic fracturing is
\begin{equation}
    P_f > -\sigma_{yy} + \sigma_T.
\end{equation}
Assuming a steady-state, hydraulic fracturing thus requires that the sandstone pressure $P_s$ increases faster than $-\sigma_{yy}$, 
\begin{equation}
    \pd{P_s}{t} > -\pd{\sigma_{yy}}{t}.
\end{equation}
We use this condition to predict if hydraulic fracturing will occur under different stress conditions. If we assume $\partial_t \sigma_{yy} = 0$, then vertical hydraulic fracturing will always occur. However, if we instead assume $\partial_t e_{yy} = 0$, then 
\begin{equation}
    (\lambda_s + 2\mu_s) \pd{e}{t} = \alpha_s \pd{P_s}{t} - 2\mu_s \dot e_{xx}.
\end{equation}
From the storage equation for the sandstone,
\begin{equation}
    \pd{P_s}{t} = \frac{2 \alpha_s \mu_s \dot e_{xx}}{\alpha_s^2 + S_s(\lambda_s + 2\mu_s)} - \frac{\lambda_s + 2\mu_s}{\alpha_s^2 + S_s(\lambda_s + 2\mu_s)} \frac{q}{h_s},
\end{equation}
where $q < 0$ indicates flow from the mudstone into the sandstone. From the condition for vertical hydraulic fracturing, we obtain
\begin{equation}
    2\mu_s \dot e_{xx} [\alpha_s(1-\alpha_s) - S_s\lambda_s] - [\lambda_s + 2\mu_s(1-\alpha_s)] \frac{q}{h_s} > 0.
\end{equation}
To simplify interpretation, we assume $\alpha \sim 1$, giving
\begin{equation}
    \dot e_{xx} < -\frac{q/h_s}{2\mu_s S_s}.
\end{equation}
If the mudstone is impermeable, $q = 0$, thus vertical hydraulic fracturing will never occur. Indeed, an additional flux from the mudstone is required for fracturing when $\partial_t e_{yy} = 0$. This only occurs when the sandstone is more compressible than the mudstone, which is not commonly observed. Furthermore, if the sandstone is highly compressible, i.e., $S_s$ is large, a higher flux may be required to satisfy the inequality.

However, assuming $\partial_t e_{yy} = 0$ is likely less physically realistic as pore pressure increases of, for example, $\sim$10 MPa would lead to predictions of $\sim$10 MPa differential stresses; these stresses are typically relieved by faulting in natural systems. For this reason we assume $\partial_t \sigma_{yy} = 0$ in the main text.

\section{Compression solution}
\label{sup:compsol}

The Laplace transform $\mathbb{F}$ of a function $f$ is defined as
\begin{equation}
    \mathbb{F}(s) = \int_0^\infty f(t) \exp(-st) \, \infd t.
\end{equation}
Transforming equations (\ref{eqn:nondiminternal})--(\ref{eqn:nondimbdy}) to Laplace space using $p_m^*(z^*, 0) = 0$ gives a second-order ordinary differential equation,
\begin{equation}
     \tdtwo{\mathbb{P}_m^*}{z^*} - s^*\mathbb{P}_m^* = -\frac{1}{s^*} \hspace{0.5cm} \mathrm{for} \,\, z^* \in (0, 1),
\end{equation}
with boundary conditions
\begin{equation}
    \begin{rcases}
        \mathbb{P}_m^* = \dfrac{\gamma}{{s^*}^2} + \dfrac{\nu}{s^*} \dfrac{\infd \mathbb{P}_m^*}{\infd z^*} \hspace{0.5cm} &\mathrm{at} \,\, z^*=0, \\
        \dfrac{\infd\mathbb{P}_m^*}{\infd z^*} = 0 \hspace{0.5cm} &\mathrm{at} \,\, z^*=1.
    \end{rcases}
\end{equation}
The solution to this system of equations is
\begin{equation}
\label{eqn:lapsol}
    \mathbb{P}_m^* = \frac{1}{{s^*}^2} + \frac{\gamma - 1}{{s^*}^2} \frac{\cosh (1-z^*)\sqrt{s^*}}{\cosh\sqrt{s^*} + \nu \sinh \sqrt{s^*}/\sqrt{s^*}}.
\end{equation}
The inverse Laplace transform can be expressed as
\begin{equation}
    f(t) = \frac{1}{2\pi i} \lim_{T\rightarrow\infty} \int_{G-iT}^{G+iT} \mathbb{F}(s) \exp(st) \, \infd s,
\end{equation}
where $G$ is greater than the real part of all the singularities of $\mathbb{F}(s)$. Assuming that the integrand is single-valued, it follows from the Cauchy residue theorem that the integral equals the sum of the residues at all the poles $s_1, s_2, ..., s_j$,
\begin{equation}
    f(t) = \sum_j \underset{s = s_j}{\mathrm{Res}} \Big\{ \mathbb{F}(s_j) \exp(s_j t)\Big\}.
\end{equation}
Equation~(\ref{eqn:lapsol}) has poles at $s^* = 0$ and at the zeros of the denominator of the second term, i.e.,
\begin{equation}
    \tanh \sqrt{s^*} = -\frac{\sqrt{s^*}}{\nu}.
\end{equation}
The solutions to this equation are imaginary, so by substituting $\sqrt{s^*} = i \xi$ we obtain the transcendental equation
\begin{equation}
    \tan\xi = -\frac{\xi}{\nu}.
\end{equation}
For the pole at $s^* = 0$, using Taylor expansions we obtain the late-time solution
\begin{equation}
    p_m^*(z^*, \infty) = \frac{\gamma + \nu}{1 + \nu} t^* + \frac{\gamma-1}{1 + \nu}\bigg\{ \frac{\nu}{3(1 + \nu)} - \tfrac{1}{2}z^*(2-z^*) \bigg\}.
\end{equation}
The residue of a quotient $f(s)/g(s)$ at a simple zero, $s_0$ is given by
\begin{equation}
    \underset{s=s_0}{\mathrm{Res}}(f/g) = \lim_{s\rightarrow s_0}\frac{f(s)}{g(s)} = \frac{f(s_0)}{g'(s_0)},
\end{equation}
hence the residues at $s^* = s_j^*$ can be expressed as,
\begin{equation}
    -2 \nu (\gamma - 1)\sum_{j=1}^\infty \frac{ \cos \{\xi_j (1 - z^*)\}}{\cos\xi_j}\frac{\exp(-\xi_j^2 t^*) }{\xi_j^2 (\nu^2 + \nu + \xi_j^2 )}.
\end{equation}
Combining residues gives the desired solution,
\begin{equation}
    p_m^* = \frac{\gamma + \nu}{1 + \nu} t^* + \frac{\gamma-1}{1 + \nu}\bigg\{ \frac{\nu}{3(1 + \nu)} - \tfrac{1}{2}z^*(2-z^*) \bigg\} - 2 \nu (\gamma - 1)\sum_{j=1}^\infty \frac{ \cos \{\xi_j (1 - z^*)\}}{\cos\xi_j}\frac{\exp(-\xi_j^2 t^*) }{\xi_j^2 (\nu^2 + \nu + \xi_j^2 )},
\end{equation}
which can be expressed more compactly, with
\begin{equation}
    p_m^* = t^* + (\gamma-1) \bigg\{ \frac{t^*}{1+\nu} + 2 \nu \sum_{j=1}^\infty \frac{ \cos \{\xi_j (1-z^*)\} }{\cos\xi_j}\frac{1-\exp(-\xi_j^2 t^*) }{\xi_j^2 (\nu^2 + \nu + \xi_j^2 )}\bigg\}.
\end{equation}

\subsection{Early time compression}
\label{sup:compearly}
At early times, $\exp(-\xi_j^2 t^*) \sim 1 - \xi_j^2 t^*$,
\begin{equation}
    p_m^* \sim t^* + (\gamma-1) \bigg\{ \frac{1}{1+\nu} + 2 \nu \sum_{j=1}^\infty \frac{ \cos \{\xi_j (1-z^*)\} }{\cos\xi_j}\frac{1}{\nu^2 + \nu + \xi_j^2 }\bigg\} t^*.
\end{equation}
Therefore, the mudstone overpressure initially increases linearly with time. By applying the initial condition $\tilde{p}_m^*(z^*, 0) = -H(-z^*)$, where $H$ is the Heaviside function, to the fluid venting solution (\ref{eqn:pipesol}), we find that
\begin{equation}
    \sum_{j=1}^\infty \frac{ \cos \{\xi_j (1-z^*)\} }{\cos\xi_j}\frac{1}{\nu^2 + \nu + \xi_j^2 } = \frac{(1+\nu)H(-z^*)-1}{2\nu(1+\nu)}.
\end{equation}
Therefore, 
\begin{equation}
    p_m^* \sim \{1 + (\gamma-1) H(-z^*) \} t^*,
\end{equation}
i.e., $p_s^* = p_m^*(0,t) \sim \gamma t^*$ and, for $z^* > 0$, $p_m^* \sim t^*$.

\section{Fluid venting solution}
\label{sup:ventsol}

For fluid venting, we assert that at $t^* = t_f^*$ the sandstone instantaneously drops in pressure by $\sigma_T^*$ such that
\begin{equation}
    \pd{p_m^*}{t^*}\bigg|_{z^*=0} = \gamma + \nu \pd{p_m^*}{z^*}\bigg|_{z^*=0} - \sigma_T^* \delta(t^* - t_f^*).
\end{equation}
Transforming the augmented system of equations to Laplace space using $p_m^*(z^*, 0) = 0$, we have
\begin{equation}
    \tdtwo{\mathbb{P}_m^*}{z^*} = \mathbb{P}_m^* - \frac{1}{{s^*}^2},
\end{equation}
with boundary conditions
\begin{equation}
    \begin{rcases}
        \mathbb{P}_m^* = \dfrac{\gamma}{{s^*}^2} + \dfrac{\nu}{{s^*}} \dfrac{\infd\mathbb{P}_m^*}{\infd z^*}\bigg|_{z^*=0} - \dfrac{\sigma_T^*\e^{-s^* t_f^*}}{{s^*}} \hspace{0.5cm} &\mathrm{at} \,\, z^*=0, \\
        \dfrac{\infd\mathbb{P}_m^*}{\infd z^*} = 0 \hspace{0.5cm} &\mathrm{at} \,\, z^*=1.
    \end{rcases}
\end{equation}
The particular solution to this system of equations is
\begin{equation}
\label{eqn:ventpart}
    \mathbb{P}_m^* = \frac{1}{{s^*}^2} + \frac{\gamma - 1}{{s^*}^2} \frac{\cosh (1-z^*)\sqrt{s^*}}{\cosh\sqrt{s^*} + \nu \sinh \sqrt{s^*}/\sqrt{s^*}} - \frac{\sigma_T^*\e^{-s^*t_f^*}}{{s^*}} \frac{\cosh (1-z^*)\sqrt{s^*}}{\cosh\sqrt{s^*} + \nu \sinh \sqrt{s^*}/\sqrt{s^*}}.
\end{equation}
We know the inverse transform of the first two terms from the solution for compression in the absence of venting (Supplementary Material \ref{sup:compsol}). The third term describes the system response to venting, thus we discard the former two terms for the present analysis. We denote the venting response with $\tilde{p}^*_m$. The third term of Eqn. (\ref{eqn:ventpart}) has residue at $s^* = 0$, giving the late-time solution
\begin{equation}
    \tilde{p}_m^*(z^*, \infty) = -\frac{\sigma_T^*}{1 + \nu}.
\end{equation}
Adding the residue at the poles of the denominator $s_j^*$, we get
\begin{align}
    \tilde{p}_m^* = -\frac{\sigma_T^*}{1 + \nu} - 2\nu \sigma_T^* \sum_{j=1}^\infty \frac{\cos \{\xi_j (1-z^*)\} }{\cos \xi_j } \frac{\exp\{-\xi_j^2 (t^*-t_f^*)\} }{\nu^2 + \nu + \xi_j^2 },
\end{align}
where $\xi_j$ is the $j$-th solution to $\tan \xi = - \xi/\nu$.

\section{Transition to periodic behaviour}
\label{sec:transition}

\begin{figure}[!htbp]
\centering
\begin{subfigure}[b]{0.99\textwidth}
   \includegraphics[width=1\linewidth]{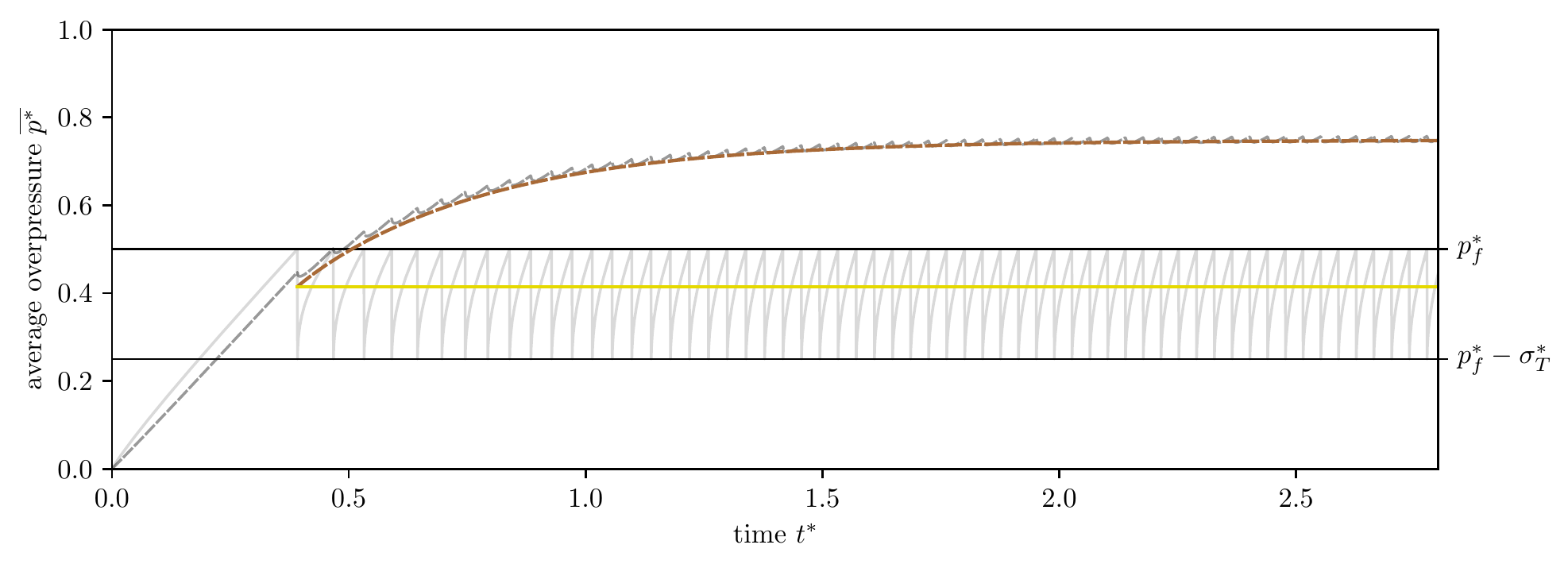}
\end{subfigure}
\caption{Average overpressure evolution of the mudstone layer (brown dashed) assuming that the sandstone layer (solid yellow) has a constant pressure. We compare this constant-pressure solution with the full solution from Fig.~\ref{fig:diff_vs_no_diff}b, plotted in grey, with $\gamma = 2$, $\nu = 5$, $p_f^* = 0.5$ and $\sigma_T^* = 0.25$. The sandstone overpressure is fixed at the time-average of the periodic venting solution, $p_s^* \approx 0.4147$. We initiate the constant-pressure solution at the time of the first venting episode $t^* \approx 0.3917$. 
\label{fig:transition}}
\end{figure}

To understand the pressure evolution without oscillations due to venting, we take the sandstone pressure to be a constant. For simplicity, we set $p_s^* = 0$ so the system of equations become
\begin{equation}
    \pd{p_m^*}{t^*} = 1 + \pdtwo{p_m^*}{z^*} \hspace{0.5cm} \mathrm{for} \,\, z^* \in [0, 1].
\end{equation}
and
\begin{equation}
    \begin{rcases}
        \displaystyle p_m^* = 0 \hspace{0.5cm} &\mathrm{at} \,\, z^*=0, \\
        \displaystyle\pd{p_m^*}{z^*} = 0 \hspace{0.5cm} &\mathrm{at} \,\, z^*=1.
    \end{rcases}
\end{equation}
Solving this system in Laplace space gives
\begin{equation}
\label{eqn:81}
    \mathbb{P}_m^* = \frac{1}{{s^*}^2} \bigg[ 1 - \frac{\cosh (1-z^*) \sqrt{s^*}}{\cosh\sqrt{s^*}} \bigg].
\end{equation}
Equation (\ref{eqn:81}) has a pole at $s^* = 0$, with residue representing the steady-state solution
\begin{equation}
    p_m^*(z^*, \infty) = -\tfrac{1}{2}z^*(2-z^*).
\end{equation}
The remainder of the poles are at the zeros of the denominator of the second term, i.e.~$s_n^* = -(2n-1)^2 \pi^2/4$ for $n \in \mathbb{Z}$. Combining residues gives the full solution,
\begin{equation}
        p_m^* = \tfrac{1}{2}z^*(2-z^*) + \frac{16}{\pi^3} \sum_{n=1}^\infty \frac{(-1)^n}{(2n-1)^3} \cos\{(1-z^*) (2n-1)\pi/2 \} \exp\{-t^* \, (2n-1)^2 \pi^2/4 \}.
\end{equation}
Therefore, the average mudstone pressure evolves such that,
\begin{equation}
    \overline{p_m^*} = \frac{1}{3} - \frac{32}{\pi^4} \sum_{n=1}^\infty \frac{\exp\{-t^* \, (2n-1)^2 \pi^2/4\}}{(2n-1)^4}.
\end{equation}
We compare the constant-pressure solution with the full solution for an example scenario in Fig.~\ref{fig:transition}. There is good agreement between the two solutions for the mudstone overpressure, with the constant-pressure solution generally following the full solution but without the oscillations produced by venting. Minor deviations between solutions at early times ($t^* \sim 0.4$) are explained by differences in their initial states.

\section{Venting period}
\label{sup:period}

After many venting episodes have occurred, the system tends towards a state of periodic venting, with a (dimensionless) time interval $\tau^*$ between episodes. At late times of tectonic compression, sandstone pressure increases linearly with time. Using the principle of superposition, we represent this system as
\begin{align}
    p_s^* = t^*\frac{\gamma + \nu}{1 + \nu} + 2\nu \sigma_T^*\sum_{n=0}^N\sum_{j=1}^\infty \frac{1 - \exp(-\xi_j^2 t^*) }{{\nu}^2 + \nu + \xi_j^2 }\exp(-\xi_j^2 n \tau^*),
\end{align}
where the number of previous venting episodes $N$ is large. The two most recent venting episodes occur at $t_{N-1}^*$ and $t_{N}^*$, such that the pressures are given by
\begin{equation}
    \begin{rcases}
        p_s^*(t_{N-1}^*) = t_{N-1}^*\dfrac{\gamma + \nu}{1 + \nu} - 2\nu \sigma_T^*\displaystyle\sum_{n=0}^{N-1}\sum_{j=1}^\infty \frac{1 - \exp(-\xi_j^2 t_{N-1}^*) }{{\nu}^2 + \nu + \xi_j^2 }\exp(-\xi_j^2 n \tau^*), \\
        p_s^*(t_{N}^*) = t_{N}^*\dfrac{\gamma + \nu}{1 + \nu} - 2\nu \sigma_T^*\displaystyle\sum_{n=0}^N\sum_{j=1}^\infty \frac{1 - \exp(-\xi_j^2 t_N^*) }{{\nu}^2 + \nu + \xi_j^2 }\exp(-\xi_j^2 n \tau^*),
    \end{rcases}
\end{equation}
We solve for $\tau^*$ using $p_s^*(t_N^*) = p_s^*(t_{N-1}^*)$. Taking $N \rightarrow \infty$ gives
\begin{equation}
    \tau^* = \frac{\sigma_T^*}{\gamma + \nu},
\end{equation}
as the dimensionless venting period. 

\section{Disequilibrium compaction}
\label{sup:diseq}

Disequilibrium compaction in this system will act to increase the total vertical stress $\sigma_{zz}$. We assume a constant rate of compressive vertical stress generation $-\dot\sigma_{zz}$ (with $\dot \sigma_{zz} \ge 0$) so from Terzaghi's principle,
\begin{equation}
    \pd{\sigma_{zz}'}{t} = -\dot\sigma_{zz} + \alpha \pd{p}{t}.
\end{equation}
Assuming sedimentation does not affect total stresses in the $x$ and $y$ directions,
\begin{equation}
    (3\lambda + 2\mu) \pd{e}{t} = -\dot\sigma_{zz} + 3 \alpha \pd{p}{t}.
\end{equation}
Using the storage equation and Darcy's law, we obtain
\begin{subequations}
\begin{align}
    \td{p_s}{t} = \frac{\alpha_s \dot \sigma_{zz}}{3\alpha_s^2 + S_s(3\lambda_s + 2\mu_s)} + \frac{3\lambda_s + 2\mu_s}{3 \alpha_s^2 + S_s(3\lambda_s + 2\mu_s)} \frac{1}{h_s}\frac{k_m}{\eta}\pd{p_m}{z}\bigg|_{z=0} \hspace{0.5cm} &\mathrm{at} \,\, z = 0, \\
    \pd{p_m}{t} = \frac{\alpha_m \dot \sigma_{zz}}{3\alpha_m^2 + S_m(3\lambda_m + 2\mu_m)} + \frac{3\lambda_m + 2\mu_m}{3\alpha_m^2 + S_m(3\lambda_m + 2\mu_m)} \frac{k_m}{\eta} \pdtwo{p_m}{z} \hspace{0.5cm} &\mathrm{for} \,\, z \in [0, h_m].
\end{align}
\end{subequations}
In a similar way as done for tectonic compression, we introduce
\begin{equation}
    D_m = \frac{k_m}{\eta} \frac{3\lambda_m + 2\mu_m}{3\alpha_m^2 + S_m(3\lambda_m + 2\mu_m)}, \hspace{1cm} D_s = \frac{k_m}{\eta}\frac{3\lambda_s + 2\mu_s}{3 \alpha_s^2 + S_s(3\lambda_s + 2\mu_s)},
\end{equation}
\begin{equation}
    \Gamma_m = \frac{\alpha_m \dot \sigma_{zz}}{3\alpha_m^2 + S_m(3\lambda_m + 2\mu_m)}, \hspace{1cm} \Gamma_s = \frac{\alpha_s \dot \sigma_{zz}}{3\alpha_s^2 + S_s(3\lambda_s + 2\mu_s)}.
\end{equation}
This selection of parameters recovers the same governing equations as for tectonic compression, Eqns. (\ref{eqn:gov1}) and (\ref{eqn:gov2}). This allows us to provide solutions to both physical problems by solving one set of equations.

\newpage
\section{Residual venting}
\label{sup:residvent}

\begin{figure}[!htbp]
\centering
\begin{subfigure}[b]{0.99\textwidth}
   \includegraphics[width=1\linewidth]{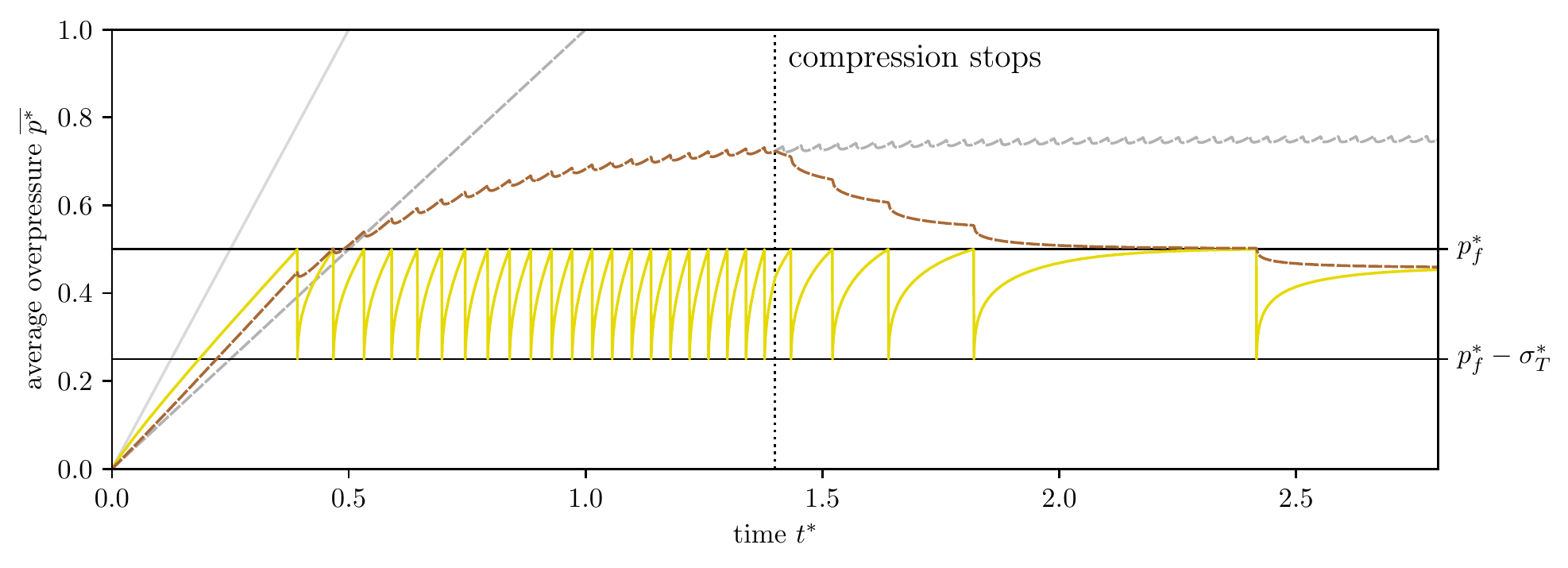}
\end{subfigure}
\caption{Average overpressure evolution of the sandstone (solid yellow) and mudstone (brown dashed) layers with compression stopping at $t^* = 1.4$. The solution for the mudstone with ongoing compression is plotted in grey. This example scenario follows Fig.~\ref{fig:diff_vs_no_diff}b, with $\gamma = 2$, $\nu = 5$, $p_f^* = 0.5$ and $\sigma_T^* = 0.25$.
\label{fig:compstop}}
\end{figure}

If compression suddenly stops, venting may continue due to pressure diffusion. We take the example presented in Fig.~\ref{fig:diff_vs_no_diff}b, and assert that compression stops at $t^* = 1.4$, plotted in Fig.~\ref{fig:compstop}. Five further venting episodes are observed before the mudstone and sandstone layers reach pressure equilibrium. The time interval between residual venting episodes increases exponentially as the mudstone pressure approaches $p_f^*$. The number of residual venting episodes will decrease by increasing the pressure drop in the mudstone per episode. This is achieved by increasing the tensile strength of the impermeable layer $\sigma_T^*$ or decreasing the capacitance ratio $\nu$. In some cases, no residual episodes are observed. 

\end{document}